\documentclass[11pt,a4paper]{article}

\usepackage{jheppub}
\usepackage{epsfig}

\usepackage{amssymb,amsmath}
\usepackage{amsthm}
\usepackage{amsbsy}

\title{Strong-coupling phases of 3D Dirac and Weyl semimetals. A 
renormalization group approach.}

\author{J. Gonz\'alez}

\affiliation{Instituto de Estructura de la Materia,\\
        Consejo Superior de Investigaciones Cient\'{\i}ficas,\\ Serrano 123,
        28006 Madrid, Spain}

\emailAdd{gonzalez@iem.cfmac.csic.es}

\abstract{We investigate the strong-coupling phases that may arise in 3D Dirac 
and Weyl semimetals under the effect of the long-range Coulomb interaction, 
considering the many-body theory of these electron systems as a variant of the
conventional fully relativistic Quantum Electrodynamics (QED). For this purpose, 
we apply two different nonperturbative approaches, consisting in the sum of 
ladder diagrams and taking the limit of a large number $N$ of fermion 
flavors. We benefit from the renormalizability that the theory shows 
in both cases to compute the anomalous scaling dimensions of different 
operators exclusively in terms of the renormalized coupling constant, allowing 
us to determine the precise location of the singularities signaling the onset 
of the strong-coupling phases. We show then that the QED of 3D Dirac semimetals 
has two competing effects at strong coupling. One of them is the tendency to 
chiral symmetry breaking and dynamical mass generation, which are analogous to 
the same phenomena arising in the conventional QED at strong coupling. This 
trend is however outweighed by the strong suppression of electron 
quasiparticles that takes place at large $N$, leading to a different type 
of critical point at sufficiently large interaction strength, shared also by 
the 3D Weyl semimetals. Overall, the phase diagram of the 3D Dirac semimetals 
turns out to be richer than that of their 2D counterparts, displaying a 
transition to a phase with non-Fermi liquid behavior which may be observed in 
materials hosting a sufficiently large number of Dirac or Weyl points.}


\notoc
\begin{document}

\maketitle

\section{Introduction}

In recent years, condensed matter physics has witnessed several remarkable 
discoveries that, starting from graphene\cite{novo}, have shown the existence 
of electron systems where the quasiparticles have linear dispersion about 
degenerate points at the intersection between valence and conduction bands. 
This has led to introduce the class of so-called Dirac semimetals, in which 
the low-energy excitations can be described in terms of a number of Dirac 
spinors. Thus, there is already clear evidence that materials like Na$_3$Bi 
or Cd$_3$As$_2$ have Dirac points in their electronic 
dispersion\cite{liu,neu,bor}, providing in that respect a 3D electronic 
analogue of graphene. More recently, there have been claims that materials 
like TaAs or the pyrochlore iridates should be examples of 3D Weyl 
semimetals\cite{taas1,taas2}, characterized by having Weyl points hosting each 
of them fermions with a given chirality.

Since the appearance of graphene, the possibility that the quasiparticles of a
condensed matter system may behave as Dirac fermions has been very appealing,
as the particular geometrical features of these mathematical objects have
shown to be indeed at the origin of several unconventional effects. The Klein 
paradox\cite{kats} and the difficulty that such quasiparticles have to undergo 
backscattering\cite{ando} are a direct consequence of the additional pseudospin 
degree of freedom inherent to Dirac fermions. The topological insulators and 
their peculiar surface states can be taken also as an illustration of the 
fruitful interplay arising between the symmetry properties of Dirac fermions 
and the topology of the space\cite{kane,qi}. 

Another important feature of known Dirac semimetals is that they need to be 
modeled as interacting electron systems which are naturally placed in the 
strong-coupling regime. This is so as the long-range Coulomb repulsion becomes
the dominant interaction, while the Fermi velocity of the electrons 
takes values which are at least two orders of magnitude below the speed of 
light. Thus, the equivalent of the fine structure constant in the electron 
system can be more than one hundred times larger than its counterpart in the 
conventional fully relativistic Quantum Electrodynamics (QED). This leads to 
expect relevant many-body effects as, for instance, the imperfect screening of 
impurities with sufficiently large charge on top of 
graphene\cite{nil,fog,shy,ter}. Another important property is the scale 
dependence of physical observables, like that predicted theoretically for 
the Fermi velocity in 2D Dirac semimetals\cite{np2,prbr} and measured recently 
in suspended graphene samples at very low doping levels\cite{exp2}. We may 
think therefore of the Dirac semimetals as an ideal playground to observe 
scaling phenomena that would be otherwise confined to the investigation of 
field theories in particle physics. 

In any event, a proper account of the scaling effects must rely on the 
renormalizability of the many-body theory. The formulation of the interacting
theory of Dirac semimetals requires the introduction of a high-energy cutoff 
in the electronic spectrum, which can be set to a rather arbitrary level. In 
order to ensure the predictability of the theory, it becomes therefore crucial 
that all the dependences on the high-energy cutoff can be absorbed into the 
redefinition of a finite number of parameters. In the case of Dirac semimetals, 
such a renormalizability cannot be taken for granted, as the many-body theory 
does not enjoy the full covariance which enforces that property in typical 
relativistic field theories. Nevertheless, the investigation of the 2D Dirac 
semimetals has already provided evidence that the interacting field theory in 
these electron systems is renormalizable\cite{jhep}. This condition has allowed 
for instance to compute very high order contributions to the renormalization of 
relevant order parameters, leading to an estimate of the critical coupling for 
chiral symmetry breaking that agrees with the value obtained with a quite 
different nonperturbative method like the resolution of the gap 
equation\cite{gama}.

The aim of this paper is to investigate the strong-coupling phases that may 
arise in 3D Dirac and Weyl semimetals under the effect of the long-range 
Coulomb interaction. In this respect, the many-body theory of these electron
systems can be viewed as a variant of conventional QED. As already pointed out,
a main difference with this theory lies in that the Fermi velocity of the 
electrons does not coincide with the speed of light, which makes that 
quasiparticle parameter susceptible of being renormalized and therefore 
dependent on the energy scale. In three spatial dimensions, the electron charge 
needs also to be redefined to account for cutoff dependences of the bare 
theory, which renders the question of the renormalizability even more 
interesting in the present context. 

In this paper we will apply two different nonperturbative approximations to 
the QED of 3D Dirac and Weyl semimetals, namely the ladder approach and the 
large-$N$ approximation ($N$ being the number of different fermion flavors), 
which can be viewed as complementary computational methods. In both cases, we 
will see that the field theory appears to be renormalizable, making possible 
to absorb all the dependences on the high-energy cutoff into a finite number 
of renormalization factors given only in terms of the renormalized coupling.
  
We will show that the QED of 3D Dirac semimetals has two competing effects at 
strong coupling. One of them is the tendency to chiral symmetry breaking and 
dynamical mass generation, which are analogous to the same phenomena arising 
in the conventional QED at strong 
coupling\cite{mas,fom,fuk,mir,gus2,kon,atk,min}. This trend is however 
outweighed by the strong suppression of electron quasiparticles that takes 
place at large $N$, leading then to a different type of critical point at 
sufficiently large interaction strength\cite{rc}, shared also by the 3D Weyl 
semimetals. We will see that the nonperturbative approaches applied in the 
paper afford a precise characterization of the respective critical behaviors 
in terms of the anomalous dimensions of relevant operators. At the end, the 
phase diagram of the 3D Dirac semimetals turns out to be richer than that of 
their 2D counterparts, displaying a transition to a phase with non-Fermi liquid 
behavior (genuine also of the 3D Weyl semimetals) which may be observed in 
materials hosting a sufficiently large number of Dirac or Weyl points.

\section{Field theory of 3D Dirac and Weyl semimetals}

Our starting point is the field theory describing the 3D fermions with linear 
dependence of the energy $\varepsilon $ on momentum ${\bf p}$, 
$\varepsilon ({\bf p}) = v_F |{\bf p}|$, and interacting through the scalar 
part $\phi $ of the electromagnetic potential. The fermion modes can be 
encoded into a number $N$ of spinor fields $\psi_i$, $i = 1, \ldots N$, where 
the index $i$ represents in general the way in which the modes are
distributed into different Dirac or Weyl points in momentum space. For 
convenience, we will choose to pair the fermion chiralities into four-component 
spinors, extending the notation to think of each $\psi_i$ field as being made 
of two different chiralities from a given Dirac point, or from two different 
Weyl points in the case of a Weyl semimetal. The hamiltonian can be written 
then as
\begin{equation}
H = - i v_F \int d^3 r \; \overline{\psi}_i({\bf r}) 
 \boldsymbol{\gamma}   \cdot \boldsymbol{\partial}  \psi_i ({\bf r})  +
  e_0  \int d^3 r \; \overline{\psi}_i({\bf r})  \gamma_0 
    \psi_i ({\bf r})  \:  \phi ({\bf r}) 
\label{h} 
\end{equation}
where $\overline{\psi}_i = \psi_i^{\dagger} \gamma_0 $ and $\{ \gamma_{\sigma} \}$
is a collection of four-dimensional matrices such that 
$\{ \gamma_\mu, \gamma_\nu \} = 2 \: {\rm diag } (1,-1,-1,-1)$.

In order to complete the field theory, we have to provide the propagator of the
scalar field $\phi $, for which we take the Lorentz gauge in the dynamics of 
the full electromagnetic potential. Thus, we get from the original relativistic 
theory
\begin{equation}
\langle T \phi ({\bf r}, t) \phi ({\bf r}', t') \rangle  =
 -i \int \frac{d^3 q}{(2\pi )^3}  \int \frac{d \omega}{2\pi } 
   e^{i {\bf q} \cdot ({\bf r} - {\bf r}')} e^{-i \omega (t-t')}
   \frac{1}{{\bf q}^2  - \omega^2/c^2 - i \eta }
\label{lorentz}
\end{equation}
We are interested in systems where the Fermi velocity $v_F$ is much smaller 
than the speed of light $c$, which makes possible to adopt a simpler 
description taking the limit $c \rightarrow \infty $. This justifies that we 
can neglect magnetic interactions between our nonrelativistic fermions, and 
that we can just deal with the zero-frequency limit of (\ref{lorentz}). 
Thus, the free propagator for the scalar potential will be taken in what 
follows as 
\begin{equation}
D_0 ({\bf q}) = \frac{1}{{\bf q}^2}
\label{scalar}
\end{equation}

A remarkable property is that the hamiltonian (\ref{h}) together with the 
interaction mediated by (\ref{scalar}) define an interacting field theory that
is scale invariant at the classical level. That is, under a change in the scale
of the space and time variables
\begin{equation}
t \rightarrow  s t   \;\;\;\; ,   \;\;\;\;   {\bf r} \rightarrow  s {\bf r}   
\end{equation}
the fields can be taken to transform accordingly as
\begin{equation}
\phi ({\bf r}) \rightarrow  \frac{1}{s} \phi ({\bf r})  \;\;\;\; ,   
     \;\;\;\;   \psi ({\bf r}) \rightarrow  \frac{1}{s^{3/2}}  \psi ({\bf r})  
\end{equation}
This gives rise to a homogeneous scaling of the hamiltonian (\ref{h}), 
consistent with the transformation of an energy variable.

Such a scale invariance has important consequences, since it makes possible 
to give meaning to the divergences that the field theory develops at high 
energies. The computation of the quantum corrections requires indeed the 
introduction of a high-energy cutoff that spoils the classical scale 
invariance. However, provided the field theory is renormalizable, the 
singular dependences on the cutoff can be still absorbed into a finite number 
of renormalization factors redefining the parameters of the theory. Writing 
the action $S$ corresponding to the hamiltonian (\ref{h}), we may start with 
a formulation introducing suitable renormalization
factors $Z_{\psi }, Z_v, Z_e$ and $Z_{\phi }$:
\begin{equation}
S =  \int dt \: d^3 r \;  Z_{\psi } \: \overline{\psi}_i({\bf r}) 
 (i \gamma_0 \partial_t + i Z_v v_R \boldsymbol{\gamma}   \cdot \boldsymbol{\partial} )
            \psi_i ({\bf r})  -
  Z_e e \int dt \: d^3 r \; Z_{\psi } \: \overline{\psi}_i({\bf r})  \gamma_0 
    \psi_i ({\bf r})  \:  Z_{\phi }  \phi ({\bf r}) 
\label{s} 
\end{equation}
The gauge invariance of the model implies that $Z_e Z_{\phi } = 1$. The point
is that, if the field theory is well-behaved, one has to be able to render all 
the physical observables cutoff-independent, when written in terms of the 
renormalized parameters $v_R$ and $e$.

\begin{figure}

\vspace{0.5cm}

\begin{center}
\mbox{\epsfxsize 12cm \epsfbox{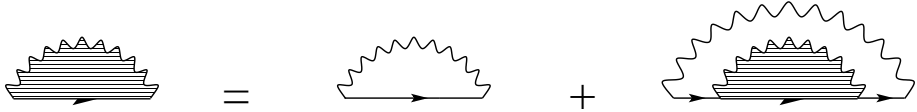}} 

\end{center}
\caption{Diagrammatic representation of the ladder approximation for the 
electron self-energy.}
\label{one}
\end{figure}

The first example of the need to implement a high-energy cutoff can be taken 
from the computation of the electron self-energy. The first perturbative 
contribution is given by the first rainbow diagram at the right-hand-side in 
Fig. \ref{one}, which corresponds to the expression
\begin{equation}
i \Sigma_1 ({\bf k}) = - e_0^2
     \int \frac{d^3 p}{(2\pi)^3}  \int \frac{d\omega_p}{2\pi }  \gamma_0
   \frac{-\gamma_0 \omega_p  + v_F \boldsymbol{\gamma} \cdot {\bf p} }
                      {-\omega_p^ 2  + v_F^2 {\bf p}^2 - i\eta }  \gamma_0 
          \frac{1}{({\bf k}-{\bf p})^2}
\label{fo}
\end{equation}
This contribution to the electron self-energy does not depend on frequency, 
but shows a divergent behavior at the upper end of the momentum integration, 
pointing at the renormalization of the Fermi velocity $v_F$. In what follows, 
instead of dealing with a cutoff in momentum space, we will prefer to use a 
regularization method that is able to preserve the gauge invariance of the 
model. For this purpose, we will adopt the analytic continuation of the 
momentum integrals to dimension $D = 3 - \epsilon $, which will convert all 
the high-energy divergences into poles in the $\epsilon $ parameter\cite{ram}. 

In the dimensional regularization method, the bare charge $e_0$ must get 
dimensions through an auxiliary momentum scale $\mu $, being related to the 
physical dimensionless charge $e$ by 
\begin{equation}
e_0 = \mu^{\epsilon/2} e
\end{equation}
The computation of the first-order contribution to the self-energy (\ref{fo})
gives then
\begin{eqnarray}
 \Sigma_1 ({\bf k})   &  =  &    \frac{e^2}{2}  \mu^\epsilon 
   \int \frac{d^D p}{(2\pi)^D}  \boldsymbol{\gamma} \cdot {\bf p} 
    \frac{1}{|{\bf p}|}  \frac{1}{({\bf k}-{\bf p})^2}        \nonumber   \\
   &  =  &  \frac{e^2}{4} \mu^\epsilon   
       \int \frac{d^D p}{(2\pi)^D}   \boldsymbol{\gamma} \cdot {\bf p} 
  \int_0^1 dx  \frac{ (1-x)^{-1/2}}{[({\bf k}-{\bf p})^2 x + {\bf p}^2 (1-x)]^{3/2}}
                                                               \nonumber        \\
   &  =  &  \frac{e^2}{4}    
       \boldsymbol{\gamma} \cdot {\bf k} \: \mu^\epsilon \int \frac{d^D p}{(2\pi)^D}    
  \int_0^1 dx  \frac{x (1-x)^{-1/2}}{[{\bf p}^2  + {\bf k}^2 x(1-x)]^{3/2}}
                                                               \nonumber        \\
    &  =  &  \frac{e^2}{2 \sqrt{\pi}}    \boldsymbol{\gamma} \cdot {\bf k} 
      \frac{\Gamma \left(\tfrac{3}{2} - \tfrac{D}{2} \right)}{(4\pi)^{D/2}} \mu^\epsilon \int_0^1 dx   
                \frac{x (1-x)^{-1/2}}{ [{\bf k}^2 x(1-x)]^{3/2 - D/2} }             \nonumber       \\    
    &  =  &   \frac{e^2}{(4\pi)^{2-\epsilon/2}}    \boldsymbol{\gamma} \cdot {\bf k} 
         \frac{\mu^\epsilon}{|{\bf k}|^{\epsilon}}
         \frac{\Gamma \left(\tfrac{1}{2}\epsilon\right) 
      \Gamma \left(2 - \tfrac{\epsilon}{2}\right) 
              \Gamma \left(\tfrac{1}{2}-\tfrac{\epsilon}{2}\right)}
                    {\Gamma \left(\tfrac{5}{2}-\epsilon \right)}
\label{loop}
\end{eqnarray}
  
The expression (\ref{loop}) shows the development of a $1/\epsilon$ pole 
as $\epsilon \rightarrow 0$. This divergence can be indeed 
absorbed into a renormalization of the Fermi velocity $v_F$, as the full 
fermion propagator $G ({\bf k}, \omega )$ is related to the self-energy
$\Sigma ({\bf k}, \omega)$ by 
\begin{equation}
G ({\bf k}, \omega )^{-1} = 
  Z_{\psi} (\gamma_0 \omega  - Z_v v_R \boldsymbol{\gamma} \cdot {\bf k}) 
      - Z_{\psi} \Sigma ({\bf k}, \omega)
\label{g_1}
\end{equation}
Thus, to first order in $e^2$, the fermion propagator can be made finite in the 
limit $\epsilon \rightarrow 0$ by taking 
\begin{equation}
Z_v = 1 - \frac{e^2}{6\pi^2 v_R}  \frac{1}{\epsilon }
\label{zv}
\end{equation}

This renormalization of the fermion propagator already gives an
idea of the effective dependence of the Fermi velocity on the energy scale. The 
bare Fermi velocity $Z_v v_R$ cannot depend on the momentum scale $\mu $, since
the original theory does not know about that auxiliary scale. We have therefore
\begin{equation}
\mu \frac{\partial }{\partial \mu } (Z_v v_R) = 0
\label{sv0}
\end{equation}
For the same reason, the bare coupling $e_0$ cannot depend on $\mu $, which 
leads to the equation
\begin{equation}
\mu  \frac{\partial }{\partial \mu } e =  - \frac{\epsilon }{2} e  
\label{se0}
\end{equation}
Combining (\ref{sv0}) and (\ref{se0}), we arrive at the scaling equation
\begin{equation}
\mu  \frac{\partial }{\partial \mu } v_R = - \frac{1}{6\pi^2 } e^2 
\label{se1}
\end{equation}
This is the expression of the growth of the renormalized Fermi velocity $v_R$ 
in the low-energy limit, approached here as $\mu \rightarrow 0$. This scaling,
which parallels the behavior of the Fermi velocity in 2D Dirac semimetals, is 
the main physical property deriving from the lowest-order renormalization of 
the 3D Dirac and Weyl semimetals\cite{ros}. 

To close this section, we comment that the result (\ref{loop}) for the electron
self-energy has actually a wider range of validity, beyond the lowest-order 
approximation in which it has been obtained. This can be seen by exploiting
a feature that is also shared with the 2D Dirac semimetals in the so-called 
ladder approximation, that is when corrections to the interaction vertex and 
the interaction propagator are neglected in the electron 
self-energy\cite{jhep}. Then we 
remain with the sum of diagrams encoded in the self-consistent equation 
represented in Fig. \ref{one}. The electron 
self-energy in this approximation, $\Sigma_{\rm ladder} ({\bf k})$, must have 
the form
\begin{equation}
\Sigma_{\rm ladder} ({\bf k}) =  f ({\bf k}) \: \boldsymbol{\gamma} \cdot {\bf k}
\end{equation}
and therefore it is bound to satisfy the equation
\begin{equation}
i\Sigma_{\rm ladder} ({\bf k}) =  i\Sigma_1 ({\bf k})    
        +  e_0^2   \int \frac{d^D p}{(2\pi)^D}   
          \int \frac{d\omega_p}{2\pi }     
\Sigma_{\rm ladder} ({\bf p}) 
    \frac{ \omega_p^2  + v_F^2 {\bf p}^2 }
                      {(-\omega_p^ 2  + v_F^2 {\bf p}^2 - i\eta)^2} 
  \frac{1}{({\bf k}-{\bf p})^2}              
\label{selfl}
\end{equation}
It is now easy to see that the second term at the right-hand-side of 
(\ref{selfl}) identically vanishes. By performing a Wick rotation 
$\omega_p = i\overline{\omega }_p$, we get 
\begin{equation}
     \int \frac{d\omega_p}{2\pi }  
        \frac{ \omega_p^2  + v_F^2 {\bf p}^2 }
                      {(-\omega_p^ 2  + v_F^2 {\bf p}^2 - i\eta)^2} 
  =  i \int \frac{d \overline{\omega}_p}{2\pi }  
    \frac{-\overline{\omega}_p^2  + v_F^2  {\bf p}^2 }
                      {(\overline{\omega }_p^ 2  + v_F^2 {\bf p}^2  )^2}
  =  0
\label{res}
\end{equation}
showing that 
\begin{equation}
\Sigma_{\rm ladder} ({\bf k})     =   \Sigma_1 ({\bf k})
\label{ident}
\end{equation}

The result (\ref{ident}) has important consequences for the ladder 
approximation, since it implies that the low-energy scaling of the Fermi 
velocity is again the main physical effect derived from the electron 
self-energy, limiting the possible electronic instabilities that 
can arise in that nonperturbative approach.

\section{3D Dirac semimetals in ladder approximation}

The iterative sequence of interactions, illustrated in Fig. \ref{one} for 
the electron self-energy, defines a partial sum of perturbation theory which 
can be also applied to analyze the renormalization of composite operators. 
This provides a suitable approach to investigate possible instabilities of the 
electron system, since some of those operators correspond to order parameters 
characterizing the breakdown of respective symmetries of the field 
theory. Once the renormalization program is accomplished, the singularities
found in the correlators of the composite operators may signal the 
condensation of a given order parameter, pointing at the onset of a new
phase of the electron system.

The most relevant composite operators are made of bilinears of the fermion
fields, having associated vertex functions that can be used to measure 
different susceptibilities of the electron system. At this point, the 
discussion is confined to the 3D Dirac semimetals (excluding the case of Weyl 
semimetals), for which we can write different composite operators in terms of 
four-component spinors about the same Dirac point in momentum space. We pay 
attention in particular to those operators corresponding to the charge, the 
current, and the fermion mass, given by
\begin{eqnarray}
\rho_0 ({\bf r}) & = & \overline{\psi}_i ({\bf r}) \gamma_0 \psi_i ({\bf r})  \\
\boldsymbol{\rho}_{c} ({\bf r}) & = &  
     \overline{\psi}_i ({\bf r}) \: \boldsymbol{\gamma} \: \psi_i ({\bf r}) \\
\rho_m ({\bf r}) & = &  \overline{\psi}_i ({\bf r})  \psi_i ({\bf r})    
\label{mass}  
\end{eqnarray}
The respective one-particle-irreducible (1PI) vertices are 
\begin{eqnarray}
 \Gamma_0 ({\bf q},\omega_q;{\bf k},\omega_k)  & = &
   \langle  \rho_0 ({\bf q},\omega_q) 
        \psi_i ({\bf k}+{\bf q},\omega_k + \omega_q) 
           \overline{\psi}_i ({\bf k},\omega_k) \rangle_{\rm 1PI}  \label{f} \\
 \boldsymbol{\Gamma}_{c} ({\bf q},\omega_q;{\bf k},\omega_k)  & = &
   \langle  \boldsymbol{\rho}_{c} ({\bf q},\omega_q) 
        \psi_i ({\bf k}+{\bf q},\omega_k + \omega_q) 
           \overline{\psi}_i ({\bf k},\omega_k) \rangle_{\rm 1PI}          \\
 \Gamma_m ({\bf q},\omega_q;{\bf k},\omega_k)  & = &
   \langle  \rho_m ({\bf q},\omega_q) 
        \psi_i ({\bf k}+{\bf q},\omega_k + \omega_q) 
           \overline{\psi}_i ({\bf k},\omega_k) \rangle_{\rm 1PI}  \label{l}
\end{eqnarray}

\begin{figure}

\vspace{0.5cm}

\begin{center}
\mbox{\epsfxsize 10cm \epsfbox{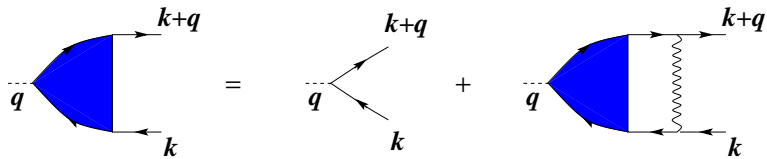}} 

\end{center}
\caption{Self-consistent diagrammatic equation for a generic vertex $\Gamma_i$
in the ladder approximation.}
\label{two}
\end{figure}

The point is that, assuming the renormalizability of the field theory, there 
must exist a multiplicative renormalization of the vertices 
\begin{eqnarray}
\Gamma_{0,{\rm ren}} & = &   Z_0 \Gamma_0         \\
\boldsymbol{\Gamma}_{c,{\rm ren}} & = &   Z_c \boldsymbol{\Gamma}_{c}     \\
\Gamma_{m,{\rm ren}} & = &   Z_m \Gamma_m        
\label{zm}
\end{eqnarray}
that renders $\Gamma_{0,{\rm ren}}, \boldsymbol{\Gamma}_{c,{\rm ren}}$ and 
$\Gamma_{m,{\rm ren}}$ independent of the high-energy cutoff. We check next
that property of the field theory, dealing with the same iteration of the 
interaction applied before to the electron self-energy. In the present context, 
this defines the ladder approximation as given by the self-consistent 
diagrammatic equation represented for a generic vertex in Fig. \ref{two}.

We first observe that the renormalization of $\Gamma_0$ must be dictated by the
gauge invariance of the model, since the composite operator $\rho_0 ({\bf r})$
already appears in the action (\ref{s}). This implies the result
\begin{equation}
Z_0 = Z_{\psi }
\label{w1}
\end{equation}
meaning that, according to the previous analysis of the self-energy, it must
be $Z_0 = 1$ in the ladder approximation. Moreover, the composite operator 
$\boldsymbol{\rho}_{c}$ may be introduced in the action multiplying it by 
the vector potential, showing that its renormalization can be related 
by gauge invariance to that of the kinetic term in (\ref{h}). We conclude
therefore that 
\begin{equation}
Z_c =  Z_{\psi }  Z_v
\label{w2}
\end{equation}
which, in the ladder approximation, leads to $Z_c = Z_v$.

The results (\ref{w1}) and (\ref{w2}) can be also obtained more formally from 
the Ward idendities that reflect the gauge invariance of the model at the 
quantum level. They lead in particular to the equations
\begin{eqnarray}
\frac{\partial }{\partial \omega} 
   \left( \gamma_0  \omega - \Sigma ({\bf k},\omega)  \right)  & = & 
   \Gamma_0 ({\bf 0},0;{\bf k},\omega)                                 \\
\frac{1}{v_F} \frac{\partial }{\partial {\bf k}} 
   \left( v_F  \boldsymbol{\gamma} \cdot {\bf k} + 
          \Sigma ({\bf k},\omega)  \right)  & = &
               \boldsymbol{\Gamma}_{c}({\bf 0},0;{\bf k},\omega)    
\end{eqnarray}
These identities were used in Ref. \cite{jhep} to check the suitability
of the dimensional regularization method to preserve the gauge invariance of 
the field theory of 2D Dirac semimetals. It was shown there that, if the 
the electron self-energy is computed with the set of diagrams encoded in the 
equation of Fig. \ref{one}, then the Ward identities are satisfied when one 
takes for the vertex the series of ladder diagrams, but dressed with the 
lowest-order electron self-energy correction. With this scope of the 
nonperturbative approach, it can be also shown by direct renormalization of 
$\boldsymbol{\Gamma}_{c}$ that $Z_c$ becomes as simple as the expression given 
by (\ref{zv}), which arises then from a remarkable cancellation of 
different contributions in the computation of the vertex.

We conclude that the charge operator is not renormalized, while the current 
operator requires a multiplicative redefinition by the simple factor (\ref{zv}) 
in the ladder approximation. The absence of a singularity for any value of the 
coupling constant excludes the possibility of having an instability related to 
the condensation of the charge or the current in the electron system. We turn 
now to the case of the mass vertex (\ref{l}), whose renormalization is not 
dictated by the previous analysis of the electron self-energy.

\subsection{Mass vertex and dynamical mass generation}

Our starting point for the analysis of the mass vertex in the ladder 
approximation is the self-consistent equation represented in 
Fig. \ref{two}. We are going to be mainly interested in the renormalization
of $\Gamma_m $ and, for that purpose, it is enough to 
consider the limit of momentum transfer ${\bf q} \rightarrow 0$ and 
$\omega_q \rightarrow 0$. The vertex must satisfy then the equation
\begin{eqnarray}
\lefteqn{\Gamma_m ({\bf 0},0;{\bf k},\omega_k) = }
                                                    \nonumber       \\
  &  &     1     +  
    i e^2_0 \int \frac{d^D p}{(2\pi )^D} \frac{d\omega_p}{2\pi } 
     \gamma_0 \frac{-\gamma_0 \omega_p  + v_F \boldsymbol{\gamma} \cdot {\bf p} }
                 {-\omega_p^2  + v_F^2 {\bf p}^2 - i\eta } 
     \Gamma_m ({\bf 0},0;{\bf p},\omega_p) 
      \frac{-\gamma_0 \omega_p  + v_F \boldsymbol{\gamma} \cdot {\bf p} }
                 {-\omega_p^2  + v_F^2 {\bf p}^2 - i\eta } \gamma_0
               \frac{1}{({\bf k}-{\bf p})^2}   \;\;\;\;\;\;\;\;\;\;
\label{selfm}
\end{eqnarray}

The resolution of (\ref{selfm}) implies that $\Gamma_m$ has to be proportional 
to the identity matrix. Moreover, it turns out to be a function independent of 
the frequency $\omega_k$ in this approximation. After a little algebra, we
arrive then at the simplified equation
\begin{equation}
\Gamma_m ({\bf 0},0;{\bf k},\omega_k) = 1  +  \frac{e_0^2}{2} 
   \int \frac{d^D p}{(2\pi )^D} \Gamma_m ({\bf 0},0;{\bf p},\omega_k) 
    \frac{1}{v_F |{\bf p}|} \frac{1}{({\bf k}-{\bf p})^2}
\label{selfcons}
\end{equation}

Eq. (\ref{selfcons}) can be solved by means of an iterative procedure, 
expressing the vertex as a power series in the effective coupling 
$\lambda_0 = e_0^2/4\pi v_F$,
\begin{equation}
\Gamma_m ({\bf 0},0;{\bf k},\omega_k) = 
 1 + \sum_{n=1}^{\infty} \lambda_0^n \: \Gamma_m^{(n)} ({\bf k})
\label{pow}
\end{equation}
Assuming that the term $\Gamma_m^{(n)} ({\bf k})$ is proportional to 
$1/|{\bf k}|^{n\epsilon }$, the next order can be computed consistently from
(\ref{selfcons}), taking into account that
\begin{eqnarray}
\Gamma_m^{(n+1)} ({\bf k})  &  \sim  & \frac{1}{2} \int \frac{d^D p}{(2\pi )^D}
  \frac{1}{|{\bf p}|^{1+n\epsilon } } \frac{1}{({\bf k}-{\bf p})^2}    
                                                        \nonumber            \\
  & = & \frac{(4\pi )^{\epsilon /2}}{16 \pi^{3/2}}   
  \frac{\Gamma \left(\tfrac{n+1}{2} \epsilon  \right) 
                      \Gamma \left(1 - \tfrac{(n+1)\epsilon}{2} \right) 
                                \Gamma \left(\tfrac{1-\epsilon}{2} \right) }
  {  \Gamma \left(\tfrac{1+n\epsilon}{2} \right) 
                        \Gamma \left(\tfrac{3 - (n + 2)\epsilon}{2} \right) }
 \frac{1}{|{\bf k}|^{(n+1)\epsilon}}
\end{eqnarray}
Thus, the vertex can be written as an expansion
\begin{equation}
\Gamma_m ({\bf 0},0;{\bf k},\omega_k) = 
 1 + \sum_{n=1}^{\infty} \lambda_0^n 
                           \frac{s_n }{|{\bf k}|^{n\epsilon}} 
\label{ser}
\end{equation}
where each order can be obtained from the previous one according to the 
recurrence relation
\begin{equation}
s_{n+1} = A_{n+1} (\epsilon) \: s_n
\label{rec}
\end{equation}
with
\begin{equation}
A_{n+1} (\epsilon) = 
  \frac{(4\pi )^{\epsilon /2}}{4 \sqrt{\pi }}   
  \frac{\Gamma \left(\tfrac{n+1}{2} \epsilon  \right) 
                         \Gamma \left(1 - \tfrac{(n+1)\epsilon}{2} \right) 
                                \Gamma \left(\tfrac{1-\epsilon}{2} \right) }
  {  \Gamma \left(\tfrac{1+n\epsilon}{2} \right) 
                          \Gamma \left(\tfrac{3 - (n + 2)\epsilon}{2} \right) }
\end{equation}

We observe from (\ref{ser}) and (\ref{rec}) that $\Gamma_m $ develops 
higher order poles in the $\epsilon $ parameter as one progresses in the 
perturbative expansion. At this point, one has to check the renormalizability
of the theory by ensuring that all the poles can be reabsorbed by means of 
a redefinition of the vertex like that in (\ref{zm}). In terms of the 
dimensionless coupling
\begin{equation}
\lambda = \frac{e^2}{4\pi v_F}
\end{equation}
the renormalization factor $Z_m$ must have the structure
\begin{equation}
Z_m = 1 + \sum_{i=1}^{\infty} \frac{d_i (\lambda )}{\epsilon^i}
\end{equation}
with residues $d_i$ depending only on $\lambda $.

In the present case, it may be actually seen that the renormalized vertex
$\Gamma_{m,{\rm ren}}$ can be made finite in the limit 
$\epsilon \rightarrow 0$, with an appropriate choice of the functions 
$d_i (\lambda )$. We obtain for instance for the first perturbative orders
\begin{eqnarray}
d_1 (\lambda ) & = &  - \frac{1}{\pi} \lambda - \frac{1}{2\pi^2} \: \lambda^2 
           -  \frac{1}{\pi^3} \left(\frac{2}{3}-\frac{\pi ^2}{36}\right)  \:  \lambda^3                           
        -  \frac{1}{12\pi^4} \left(15-\pi ^2\right)  \: \lambda^4  
                                                                \nonumber  \\
  &  &  - \frac{1}{400\pi^5} \left(1120-100 \pi ^2+\pi ^4\right) \: \lambda^5  
         -  \frac{1}{\pi^6}  \left(7-\frac{7 \pi^2}{9}+\frac{23 \pi ^4}{1440}\right)  \:  \lambda^6       
                                                             +  \ldots         \\
d_2 (\lambda ) & = &  \frac{1}{2\pi^2} \: \lambda^2 + \frac{1}{2\pi^3} \: \lambda^3 
                  +  \frac{1}{72\pi^4} \left(57-2 \pi ^2\right) \: \lambda^4    
         +  \frac{1}{72\pi^5} \left(114-7 \pi ^2\right) \: \lambda^5 
                                                                 \nonumber  \\
 &  & + \frac{1}{64800\pi^6} \left(236340-20100 \pi ^2+187 \pi ^4\right) \: \lambda^6  +  \ldots    
                                                                           \\
d_3 (\lambda ) & = & - \frac{1}{6\pi^3} \: \lambda^3 
      - \frac{1}{4\pi^4} \: \lambda^4 -  \frac{1}{72\pi^5} \left(33-\pi ^2\right) \: \lambda^5 
 - \frac{1}{\pi^6} \left(\frac{47}{48}-\frac{\pi ^2}{18}\right) \: \lambda^6 + \ldots 
                                                                           \\
d_4 (\lambda ) & = &   \frac{1}{24\pi^4} \: \lambda^4 
 +  \frac{1}{12\pi^5} \: \lambda^5 +  \frac{1}{432\pi^6} \left(75-2 \pi ^2\right) \: \lambda^6 
                                                     + \ldots               \\
d_5 (\lambda ) & = &  - \frac{1}{120\pi^5} \: \lambda^5 
              -  \frac{1}{48\pi^6} \: \lambda^6  + \ldots               \\
d_6 (\lambda ) & = &   \frac{1}{720\pi^6} \: \lambda^6   +  \ldots
\end{eqnarray}
An important point about the residues $d_i (\lambda )$ is that they can be 
chosen without having any dependence on the momentum ${\bf k}$ of the vertex. 
This is a signature of the renormalizability of the theory, by which we are
able to render it independent of the high-energy cutoff, redefining just a 
finite number of local operators. 

On the other hand, the renormalization factor $Z_m$ contains important 
information about the behavior of the theory in the low-energy limit. This 
stems from the anomalous scale dependence that the vertex gets as a consequence
of its renormalization\cite{amit}. The bare unrenormalized theory does not know 
about the momentum scale $\mu $, but the factor $Z_m$ lends to 
$\Gamma_{m,{\rm ren}}$ an anomalous scaling of the form
\begin{equation}
\Gamma_{m, {\rm ren}} \sim  \mu^{\gamma_m}
\end{equation}
The anomalous dimension $\gamma_m $ can be thus obtained as
\begin{equation}
\gamma_m = \frac{\mu }{Z_m}  \frac{\partial Z_m}{\partial \mu}
\label{adm}
\end{equation}

The renormalization factor $Z_m$ is given by an infinite series of poles in
the $\epsilon $ parameter, and then it is highly nontrivial that the 
computation of $\gamma_m $ from (\ref{adm}) may provide a finite result in 
the limit $\epsilon \rightarrow 0$. At this stage, the dependence of $Z_m$ 
on $\mu $ arises from the scaling (\ref{se0}), since no self-energy corrections
are taken into account yet. That leads to the equation
\begin{equation}
\left( 1 + \sum_{i=1}^{\infty} \frac{d_i (\lambda )}{\epsilon^i} \right) \gamma_m
   =  - \lambda  \sum_{i=0}^{\infty} \frac{d}{d\lambda }d_{i+1}(\lambda ) \: \frac{1}{\epsilon^i}
\label{altm}
\end{equation}
The term with no poles in the $\epsilon $ parameter already gives the result
for the anomalous dimension
\begin{equation}
\gamma_m  =  -\lambda  \frac{d }{d \lambda}d_1(\lambda )
\label{anomm}
\end{equation}
But it remains however to ensure that the rest of the poles identically cancel
in Eq. (\ref{altm}), which requires the fulfillment of the hierarchy of equations
\begin{equation}
\frac{d }{d \lambda}d_{i+1}(\lambda ) = d_i (\lambda ) \: \frac{d }{d \lambda}d_1(\lambda )
\label{rr}
\end{equation}

\begin{figure}

\vspace{0.5cm}

\begin{center}
\mbox{\epsfxsize 7cm \epsfbox{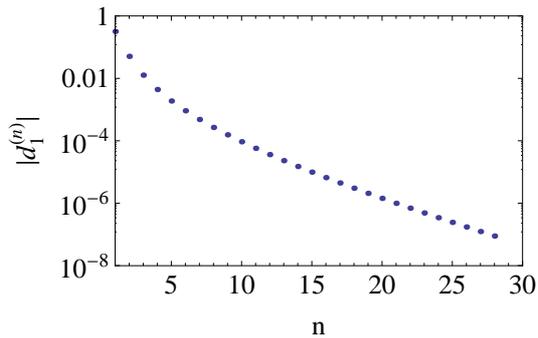}} 

\end{center}
\caption{Plot of the absolute value of the coefficients $d_1^{(n)}$ in the 
expansion of $d_1 (\lambda )$ as a power series of the renormalized coupling 
$\lambda $.}
\label{three}
\end{figure}

Quite remarkably, we have checked that the set of equations (\ref{rr}) is 
satisfied, up to the order we have been able to compute the residues 
$d_i (\lambda )$ in the ladder approximation. In this task, we have managed to 
obtain numerically the first residues up to order $\lambda^{28}$, with a 
precision of 36 digits. Besides verifying the hierarchy (\ref{rr}) at this 
level of approximation, we have also analyzed the trend of the power series for 
such large perturbative orders, finding that a singularity must exist at a 
certain critical coupling. Concentrating in particular on the function 
$d_1 (\lambda )$, we have found that the terms in its perturbative expansion
\begin{equation}
d_1 (\lambda )  = \sum_{n=1}^{\infty}  d_1^{(n)} \lambda^n
\end{equation}
approach a geometric sequence at large $n$. The values we have obtained for 
the coefficients $d_1^{(n)}$ are represented in Fig. \ref{three} up to order  
$n = 28$. From the precise numerical computation of the coefficients, it
is actually possible to verify that their ratio follows very accurately the 
asymptotic behavior 
\begin{equation}
\frac{d_1^{(n)}}{d_1^{(n-1)}} = d + \frac{d'}{n} + \frac{d''}{n^2} + \frac{d'''}{n^3} + \ldots
\label{scn}
\end{equation}
From the estimate of the limit value $d$, we have obtained the 
finite radius of convergence for a value of the coupling constant
\begin{equation}
\lambda_c = \frac{1}{d} \approx 1.27324
\label{crit}
\end{equation}

The coupling $\lambda_c $ corresponds to a point where the anomalous
dimension $\gamma_m$ diverges, which signals in turn the development of 
a singular susceptibility with respect to the mass parameter. That is,
$\lambda_c $ has to be viewed as a critical coupling above which the
electron system enters a new phase with dynamical generation of 
mass\cite{nomu}. This characterization is also supported by the fact that, in 
the case of the 2D Dirac semimetals, a similar renormalization in the ladder 
approximation\cite{jhep} has shown to lead to a value of the critical coupling 
that coincides very precisely with the point for dynamical mass generation 
determined from the resolution of the mass gap equation\cite{gama}. It is 
indeed remarkable that these quite different approaches give an identical 
result for the point of chiral symmetry breaking in the electron system. This 
reassures the predictability of our renormalization approach, that moreover 
allows to extend the analysis beyond the ladder approximation as we discuss 
in what follows.

\subsection{Electron self-energy corrections to the ladder approximation}

The most relevant way to improve the ladder approximation corresponds to 
including electron self-energy corrections in the diagrams encoded in the 
equation of Fig. \ref{two}. With these additional contributions we may account 
for an important feature of the electron system, which is the growth of the 
Fermi velocity in the low-energy limit. This leads to a reduction of the 
effective interaction strength, from which we can expect the need to push the 
nominal coupling to larger values in order to reach the phase with dynamical 
mass generation.

A fully consistent approach can be devised by considering the self-energy 
contributions in the same ladder approximation discussed in Sec. II. In this
case we know that the electron self-energy coincides with the lowest-order 
result given by Eq. (\ref{loop}). This can be translated into a redefinition 
of $v_F$, leading in the fermion propagator to an effective Fermi velocity
\begin{equation}
\tilde{v}_F ({\bf k})
    = v_F +  \frac{e_0^2}{4\pi } B(\epsilon )  \frac{1}{|{\bf k}|^{\epsilon}}
\label{til}
\end{equation}
with 
\begin{equation}
B(\epsilon ) = \frac{1}{(4\pi)^{1-\epsilon/2}}
         \frac{\Gamma \left(\tfrac{1}{2}\epsilon\right) 
      \Gamma \left(2 - \tfrac{\epsilon}{2}\right) 
              \Gamma \left(\tfrac{1}{2}-\tfrac{\epsilon}{2}\right)}
                    {\Gamma \left(\tfrac{5}{2}-\epsilon \right)}
\end{equation}
The electron self-energy corrections are then accounted for automatically after
replacing the parameter $v_F$ by $\tilde{v}_F ({\bf k})$ in the self-consistent
equation (\ref{selfcons}) for the mass vertex. It can be easily checked that 
the perturbative expansion of $1/\tilde{v}_F ({\bf k})$ introduced in that 
equation corresponds to the iteration of self-energy rainbow diagrams 
correcting the fermion propagators in the original ladder approximation.

As in the previous subsection, we can assume that the mass vertex admits now 
the expansion
\begin{equation}
\Gamma_m ({\bf 0},0;{\bf k},\omega_k) = 
 1 + \sum_{n=1}^{\infty} \lambda_0^n 
                           \frac{t_n }{|{\bf k}|^{n\epsilon}} 
\label{ser2}
\end{equation}
We can apply a recurrent procedure as before to compute the different orders in
(\ref{ser2}), with a main difference in that now each $t_n$ is going to depend 
on all the lower orders in the expansion. This is so as the $n$-th order, when 
inserted at the improved right-hand-side of (\ref{selfcons}), can give rise to
contributions to any higher order as the factor $1/\tilde{v}_F ({\bf k})$ is 
also expanded. At the end, we arrive at the result that
\begin{equation}
t_{n+1} = A_{n+1}(\epsilon) \sum_{l=0}^{n} \left(-B(\epsilon) \right)^{n-l} t_l
\end{equation}

We have to check again that the vertex can be made finite in the limit 
$\epsilon \rightarrow 0$ by a suitable multiplicative renormalization. In this
case, we have first to express all quantities in terms of the 
renormalized Fermi velocity $v_R$ arising after subtraction of the pole at the 
right-hand-side of (\ref{til}). $v_R$ is defined by 
\begin{equation}
v_F = Z_v v_R
\end{equation}
where $Z_v$ has already appeared in (\ref{zv}). The renormalized 
coupling is given now by
\begin{equation}
\lambda = \frac{e^2}{4\pi v_R}
\end{equation}
When expressed in terms of $\lambda $, the new renormalization factor $Z_m$ 
must have the structure
\begin{equation}
Z_m = 1 + \sum_{i=1}^{\infty} \frac{\tilde{d}_i (\lambda )}{\epsilon^i}
\end{equation}

The first terms in the expansions of the residues $\tilde{d}_i (\lambda )$ 
can be obtained analytically, with the result that 
\begin{eqnarray}
\tilde{d}_1 (\lambda )  & = &  - \frac{1}{\pi} \lambda - \frac{1}{9\pi^2} \: \lambda^2 
      -  \frac{76-5 \pi ^2}{324 \pi ^3}  \lambda^3               
        -   \frac{1908 -630 \zeta (3)-91 \pi ^2}{3888 \pi ^4}  \lambda^4   
                                                               \nonumber  \\
 & &  - \frac{168 (386-91 \zeta (3))-4004 \pi ^2-33 \pi ^4}{58320 \pi ^5} \: \lambda^5   
                                                                             \nonumber \\
 & &  - \frac{35 \pi ^2 (615 \zeta (3)-3014)-15 (28028 \zeta (3)+9765 \zeta (5)-94464)+632
   \pi ^4}{524880 \pi ^6}  \: \lambda^6    +    \ldots    \;\;\;\;\;   \label{df}    \\
\tilde{d}_2 (\lambda ) & = &  \frac{1}{6 \pi ^2} \: \lambda^2 
         +   \frac{5}{81 \pi ^3} \: \lambda^3                         
         +  \frac{5 \left(16-\pi ^2\right)}{648 \pi ^4} \lambda^4         
       +  \frac{14876 -4410 \zeta (3)-737 \pi ^2}{58320 \pi ^5} \: \lambda^5 
                                                              \nonumber     \\
  & &  +  \frac{604904 -141204 \zeta (3)-38562 \pi ^2-139 \pi ^4}{1049760 \pi ^6}
                 \: \lambda^6                    +  \ldots                     \\
\tilde{d}_3 (\lambda ) & = &  \frac{1}{54 \pi ^3} \: \lambda^3  
         +  \frac{1}{162 \pi ^4} \: \lambda^4 
           +   \frac{218-15 \pi ^2}{14580 \pi ^5} \lambda^5         
       +   \frac{34492 -11970 \zeta (3)-1629 \pi ^2}{1049760 \pi ^6} \:  \lambda^6
            + \ldots                                                            \\
\tilde{d}_4 (\lambda ) & = &  \frac{1}{216 \pi ^4} \: \lambda^4  
        +  \frac{1}{810 \pi ^5} \: \lambda^5   
         +  \frac{1792-135 \pi ^2}{524880 \pi ^6} \:  \lambda^6    +  \ldots        \\
\tilde{d}_5 (\lambda ) & = &   \frac{1}{648 \pi ^5} \: \lambda^5  
      +  \frac{1}{3240 \pi ^6} \: \lambda^6    +  \ldots                         \\
\tilde{d}_6 (\lambda ) & = &   \frac{7}{11664 \pi ^6} \: \lambda^6    +  \ldots      
\label{dl}
\end{eqnarray}
The important point is again that the functions $\tilde{d}_i (\lambda )$ can 
be chosen with no nonlocal dependence (in fact, with no dependence) on the 
external momentum of the vertex, which means that the renormalization can be 
accomplished with the redefinition of a finite number of purely local 
operators.

The expansion of the residue $\tilde{d}_1 (\lambda )$ allows us to estimate
the effect of the self-energy corrections on the anomalous dimension of the 
vertex. This is given as before by Eq. (\ref{adm}), while now we have to 
account for the change in the scaling of the renormalized coupling $\lambda $. 
This can be obtained from Eqs. (\ref{se0}) and (\ref{se1}), with the result 
that
\begin{equation}
\mu \frac{\partial }{\partial \mu }  \lambda = - \epsilon \lambda
   +  \frac{2}{3\pi }  \lambda^2
\end{equation}
The computation of the anomalous dimension proceeds then as 
\begin{equation}
\gamma_m =  \frac{\mu }{Z_m}  
           \frac{\partial \lambda}{\partial \mu}  
             \frac{\partial Z_m}{\partial \lambda}
\label{chr}
\end{equation}

Eq. (\ref{chr}) provides an expression for $\gamma_m $ that contains poles 
of all orders in the $\epsilon $ parameter. As in the previous subsection, it 
is therefore highly nontrivial that a finite result may be obtained for the 
anomalous dimension in the limit $\epsilon \rightarrow 0$. From (\ref{chr}), 
the equation that we have to inspect in this case is
\begin{equation}
\left( 1 + \sum_{i=1}^{\infty } \frac{\tilde{d}_i (\lambda)}{\epsilon^i} \right)  \gamma_m 
   =  - \lambda \sum_{i=0}^{\infty }  \frac{1}{\epsilon^i} \frac{d}{d\lambda } \tilde{d}_{i+1} (\lambda)
      + \frac{2}{3\pi } \lambda^2 
      \sum_{i=1}^{\infty }  \frac{1}{\epsilon^i} \frac{d}{d\lambda } \tilde{d}_i (\lambda)
\label{insp}
\end{equation}
The zeroth order in $\epsilon $ leads to the equation for the anomalous
dimension
\begin{equation}
\gamma_m = -\lambda \frac{d}{d\lambda } \tilde{d}_1 (\lambda)
\label{gamm}
\end{equation}
However, this derivation makes sense only when the cancellation of all the 
poles in (\ref{insp}) can be guaranteed, which implies the set of equations
\begin{equation}
   \frac{d}{d\lambda } \tilde{d}_{i+1} (\lambda)
 - \tilde{d}_i (\lambda) \: \frac{d}{d\lambda } \tilde{d}_1 (\lambda)  
 - \frac{2}{3\pi } \lambda \: \frac{d}{d\lambda } \tilde{d}_i (\lambda)  =  0
\label{rem}
\end{equation}

It is reassuring to see that the analytic expressions given in 
(\ref{df})-(\ref{dl}) satisfy the hierarchy of equations (\ref{rem}). A more 
comprehensive check can be performed however with the numerical computation of 
the expansions of the first residues, that we have carried out up to order 
$\lambda^{30}$ and with a precision of 40 digits. In this way, we have been 
able to certify that the conditions (\ref{rem}) are also fulfilled at that 
level of approximation.

\begin{figure}

\vspace{0.5cm}

\begin{center}
\mbox{\epsfxsize 7cm \epsfbox{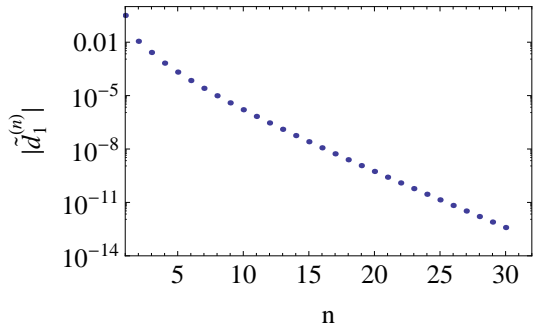}} 

\end{center}
\caption{Plot of the absolute value of the coefficients $\tilde{d}_1^{(n)}$ in the 
expansion of $\tilde{d}_1 (\lambda )$ as a power series of the renormalized 
coupling $\lambda $.}
\label{four}
\end{figure}

The residue $\tilde{d}_1 (\lambda)$ has in particular an expansion
\begin{equation}
\tilde{d}_1 (\lambda )  = \sum_{n=1}^{\infty}  \tilde{d}_1^{(n)} \lambda^n
\end{equation}
with coefficients that have been represented in Fig. \ref{four} up to order 
$n = 30$. We observe that the $\tilde{d}_1^{(n)}$ series approaches a 
geometric sequence at large $n$. The ratio between consecutive orders can be 
fitted indeed with great accuracy by a behavior like (\ref{scn}), 
allowing to compute the asymptotic value
\begin{equation}
\frac{\tilde{d}_1^{(n)}}{\tilde{d}_1^{(n-1)}}  \rightarrow  \tilde{d}      
\end{equation}
in the limit $n \rightarrow \infty$.
We have obtained in this way the finite radius of convergence of the 
perturbative series 
\begin{equation}
\lambda_c = \frac{1}{\tilde{d}}  \approx 1.8660
\label{crit2}
\end{equation}
with an error estimated to be (as in (\ref{crit})) in the last digit.

As remarked before, the critical coupling $\lambda_c$ corresponds to the point
at which the anomalous dimension $\gamma_m$ diverges. This is in turn the 
signature of the development of a nonvanishing expectation value of the mass
operator (\ref{mass}). Thus we see that, even after taking into account the 
effect of the Fermi velocity renormalization, there still remains a 
strong-coupling phase in the electron system characterized by the dynamical 
generation of mass for the Dirac fermions. The value of the critical coupling 
(\ref{crit2}) is sensibly larger than that obtained in the previous 
subsection, which is consistent with the fact that the self-energy corrections 
effectively reduce the interaction strength. The situation is in this respect 
rather similar to the case of the 2D Dirac semimetals, where the effective 
growth of the Fermi velocity at low energies has been invoked as the reason 
why no gap is observed in the electronic spectrum of graphene\cite{sabio,qed}, 
even in the free-standing samples prepared at very low doping levels about the 
Dirac point\cite{exp2}.

\section{Large-$N$ approximation for 3D Dirac and Weyl semimetals}

\begin{figure}[h]
\begin{center}
\mbox{\epsfxsize 2.0cm \epsfbox{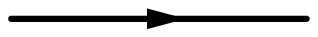}} \hspace{0.5cm} {\Large $=$}
 \hspace{0.5cm}  \mbox{\epsfxsize 2.0cm \epsfbox{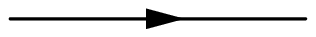}}
 \hspace{0.5cm}  {\Large $+$}  \hspace{0.5cm}
\mbox{\epsfxsize 4.8cm \epsfbox{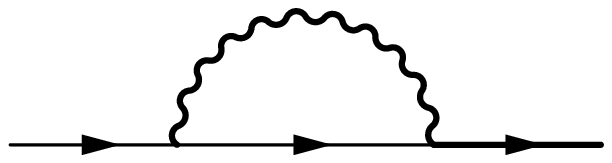}}  \\  \vspace{1cm}
% \hspace{0.3cm}  (a) \hspace{6.6cm} (b) 
\raisebox{0.2cm}{\epsfxsize 2.0cm \epsfbox{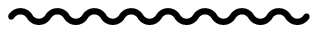}} \hspace{0.5cm} 
  \raisebox{0.16cm}{\Large $=$}
 \hspace{0.5cm}  \raisebox{0.2cm}{\epsfxsize 2.0cm \epsfbox{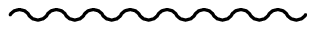}}
 \hspace{0.5cm}  
  \raisebox{0.16cm}{\Large $+$} \hspace{0.5cm}  
    \raisebox{-0.6cm}{\epsfxsize 4.8cm \epsfbox{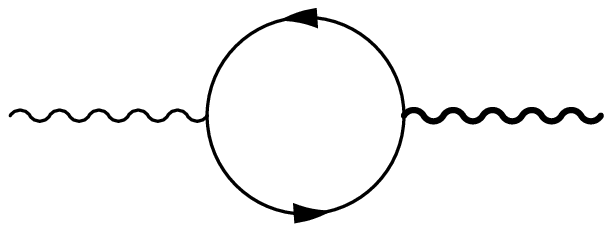}} 
\end{center}
\caption{Diagrammatic equations in the large-$N$ approximation. The thick(thin) straight line stands for the dressed(free) Dirac fermion propagator and the thick(thin) wiggly line stands for the dressed(free) interaction propagator.}
\label{five}
\end{figure}

We deal next with an approach complementary to the ladder approximation, paying
attention to the effect of the photon self-energy corrections on the electron
quasiparticles. This will take into account the renormalization of the electron
charge, which is a relevant feature in the field theory of 3D semimetals as well 
as in the fully relativistic 3D QED\cite{landau}. In order to include a 
consistent set of quantum corrections, we will dress the interaction with the 
sum of bubble diagrams obtained by iteration of the electron-hole polarization, 
as represented in Fig. \ref{five}. This approximation can be then considered as 
the result of taking the leading order in a $1/N$ expansion, providing a 
resolution of the theory in a well-defined limit as $N \rightarrow \infty$.

The corrections to the bare interaction are represented by the polarization
$\Pi ({\bf q},\omega_q )$, from which the full interaction propagator 
$D ({\bf q},\omega_q )$ is obtained according to
\begin{equation}
D ({\bf q},\omega_q )^{-1} = D_0 ({\bf q})^{-1} - \Pi ({\bf q},\omega_q )
\end{equation}
In the present approximation, the polarization is given by the dominant 
contribution in the large-$N$ limit
\begin{equation}
i\Pi ({\bf q},\omega_q ) = N e_0^2 
            \int \frac{d^D p}{(2\pi )^D} \frac{d\omega_p }{2\pi } 
{\rm Tr} \left[ \gamma_0 G_0({\bf p} + {\bf q}, \omega_p + \omega_q)   
   \gamma_0  G_0({\bf p},\omega_p ) \right]
\label{pi}
\end{equation}
with the free Dirac propagator
\begin{equation}
G_0({\bf p},\omega_p ) = 
\frac{-\gamma_0 \omega_p  + v_F \boldsymbol{\gamma} \cdot {\bf p} }
                      {-\omega_p^ 2  + v_F^2 {\bf p}^2 - i\eta }
\end{equation}
Computing the integrals in analytic continuation to $D = 3 - \epsilon $,
we get the result
\begin{equation}
\Pi ({\bf q},\omega_q ) =
- N C(\epsilon ) \frac{e_0^2}{2\pi^2 v_F} 
   \frac{{\bf q}^2}{({\bf q}^2 - \omega_q^2/v_F^2)^{\epsilon/2}}
\label{plrz}
\end{equation}
with 
\begin{equation}
C(\epsilon ) = (4\pi )^{\epsilon/2} 
     \Gamma \left( \tfrac{1}{2} \epsilon \right) 
     \frac{\Gamma \left(2 - \tfrac{\epsilon}{2} \right)^2}{\Gamma (4 - \epsilon) }
\end{equation}

The polarization (\ref{plrz}) diverges in the limit $\epsilon \rightarrow 0$,
which points at the need to renormalize the scalar field $\phi $ mediating the 
Coulomb interaction. As already mentioned, the gauge invariance of the action 
(\ref{s}) implies that $Z_e Z_{\phi } = 1$, which means that the electron 
charge is consequently renormalized. In the present approximation, these 
effects can be taken into account at once in the large-$N$ expression of the 
electron self-energy
\begin{eqnarray}
i\Sigma ({\bf k},\omega_k )  & = &  - \int \frac{d^D p}{(2\pi )^D} \frac{d\omega_p }{2\pi } \:
  \gamma_0  \frac{ -\gamma_0 (\omega_k - \omega_p ) + 
      v_F \mbox{\boldmath $\gamma $}  \cdot  ({\bf k} - {\bf p})}
   { - (\omega_k - \omega_p)^2 + v_F^2 ({\bf k} - {\bf p})^2 - i\eta}  \gamma_0  \:  
                                                                 \nonumber       \\
     &    &    \;\;\;\;\;\;\;\;\;\;\;\;\;\;\;\;\;\;\;\;\;\;\;\;\;
            \times   \frac{e_0^2}
             {{\bf p}^2 \left(1 + N C(\epsilon ) \frac{e_0^2}{2\pi^2 v_F}
       \frac{1}{({\bf p}^2 - \omega_p^2/v_F^2)^{\epsilon/2}} \right)}   
\label{selfn}  
\end{eqnarray}
Thus, we can determine the electron charge renormalization by devising
a finite limit of the effective $e$-$e$ interaction in (\ref{selfn}) as 
$\epsilon \rightarrow 0$. This can be achieved by absorbing the pole in 
(\ref{plrz}) into the bare charge $e_0$ according to the redefinition
\begin{equation}
\frac{1}{e^2} = \frac{\mu^\epsilon }{e_0^2} + \frac{N}{6\pi^2 v_F} \frac{1}{\epsilon } 
\label{eren}
\end{equation}
In terms of the renormalized charge $e$, the electron self-energy becomes
\begin{eqnarray}
i\Sigma ({\bf k},\omega_k ) & = &  \int \frac{d^D p}{(2\pi )^D} \frac{d\omega_p }{2\pi } \:
  \frac{ \gamma_0 (\omega_k - \omega_p ) + v_F \mbox{\boldmath $\gamma $}  \cdot  ({\bf k} - {\bf p})}
           { - (\omega_k - \omega_p)^2 + v_F^2 ({\bf k} - {\bf p})^2 - i\eta}  \:
                                                         \nonumber           \\
     &    &    \;\;\;\;\;\;\;\;\;\;\;\;\;\;\;\;\;\;\;\;\;\;\;\;\;   
                  \times    \frac{\mu^\epsilon e^2}
   {{\bf p}^2 \left(1 - \frac{Ne^2}{6\pi^2 v_F} \frac{1}{\epsilon }  + 
    C(\epsilon ) \frac{Ne^2}{2\pi^2 v_F}
       \frac{\mu^\epsilon}{({\bf p}^2 - \omega_p^2/v_F^2 )^{\epsilon/2}} \right)}    
\label{selfr} 
\end{eqnarray}

The renormalization program has to be completed anyhow by accounting for the
divergences that arise when performing the integrals in (\ref{selfr}). 
In order to make sense of the theory in the large-$N$ limit, we can assume 
formally that $Ne^2 / 2\pi^2 v_F \sim O(N^0 )$. The difference between $v_F$ 
and its renormalized counterpart $v_R$ is then a quantity of order $\sim 1/N$,
and the electron self-energy gets a natural dependence on the effective 
coupling
\begin{equation}
g = \frac{Ne^2}{2\pi^2 v_R}
\label{gdef}
\end{equation}
We may actually resort to a perturbative expansion
\begin{eqnarray}
\Sigma ({\bf k},i\overline{\omega}_k )  & = &  
   \frac{2\pi^2 v_F}{N}  \mu^\epsilon \int \frac{d^D p}{(2\pi )^D} \frac{d\overline{\omega}_p }{2\pi }   \:
  \frac{i \gamma_0 (\overline{\omega}_k - \overline{\omega}_p ) 
                          + v_F  \mbox{\boldmath $\gamma $}  \cdot  ({\bf k} - {\bf p})}
           { v_F^2 ({\bf k} - {\bf p})^2 + (\overline{\omega}_k - \overline{\omega}_p)^2}   \:
         \frac{1}{{\bf p}^2}                             \nonumber       \\
    &   &      \;\;\;\;\;\;\;\;\;\;\;\;\;\;\;\;\;\;\;\;\;\;\;\;\; 
      \sum_{n=0}^{\infty} (-1)^n g^{n+1} \left(- \frac{1}{3\epsilon } + C(\epsilon )
       \frac{\mu^\epsilon}{({\bf p}^2 + \overline{\omega}_p^2/v_F^2)^{\epsilon/2}} \right)^n
\label{pexp}     
\end{eqnarray}
It may be seen from (\ref{pexp}) that, to order $g^n$, the self-energy can 
develop divergences as large as $\sim 1/\epsilon^n $. If the theory is 
renormalizable, it must be possible to absorb these poles into the definition 
of suitable renormalization factors in the full fermion propagator (\ref{g_1}). 
We look then for the large-$N$ limit of $Z_\psi $ and $Z_v$, which must have 
the general structure
\begin{eqnarray}
Z_\psi  & = &  1 + \frac{1}{N}\sum_{i=1}^{\infty } \frac{c_i (g)}{\epsilon^i} 
                                      +  O \left( \frac{1}{N^2} \right)    
                                                       \label{zpn}    \\
Z_v  & = &  1 + \frac{1}{N}\sum_{i=1}^{\infty } \frac{b_i (g)}{\epsilon^i}
                                       +  O \left( \frac{1}{N^2} \right)
\label{zvn}
\end{eqnarray}
with coefficients $b_i$ and $c_i$ depending only on the renormalized coupling 
$g$.

It is not difficult to compute the integrals that are needed to get in general
the $n$-th order of $\Sigma ({\bf k},i\overline{\omega}_k )$. We are interested 
in terms that are linear in $\overline{\omega}_k$ and ${\bf k}$, and the only
technical point is that these contributions must be also regularized in the 
infrared, using for instance the own external frequency or momentum, or 
some other suitable parameter. It can be checked that the choice of a 
particular infrared regulator is not important when extracting the high-energy 
divergences as $\epsilon \rightarrow 0$. For this reason, we have resorted to 
introduce a fictitious Dirac mass $\nu $, which has the ability of keeping 
well-behaved both types of contributions linear in $\overline{\omega}_k$ and 
${\bf k}$. We rely then on the results for the generic integrals
\begin{eqnarray}
I_n  & = & v_F \int \frac{d^D p}{(2\pi )^D} \frac{d\overline{\omega}_p }{2\pi }   \:
  \frac{i \gamma_0 (\overline{\omega}_k - \overline{\omega}_p )   }
  { v_F^2 ({\bf k} - {\bf p})^2 + (\overline{\omega}_k - \overline{\omega}_p)^2 + v_F^2 \nu^2 }   \:
         \frac{1}{{\bf p}^2}  \frac{1}{({\bf p}^2 + \overline{\omega}_p^2/v_F^2)^{n\epsilon/2}}  
                                                                 \nonumber         \\
  & \approx &  i \gamma_0 \overline{\omega}_k \frac{1}{(4\pi )^{2-\epsilon/2} } 
  \frac{n\epsilon }{1-\epsilon }  
 \frac{\Gamma \left(\tfrac{n+1}{2} \epsilon  \right) 
                         \Gamma \left(1-\tfrac{n+1}{2}\epsilon \right)  }
 {\Gamma \left(2 - \tfrac{\epsilon}{2} \right)}  \frac{1}{|\nu |^{(n+1)\epsilon }} 
\label{in}
\end{eqnarray}
and
\begin{eqnarray}     
J_n  & = & v_F  \int \frac{d^D p}{(2\pi )^D} \frac{d\overline{\omega}_p }{2\pi }   \:
  \frac{ v_F  \mbox{\boldmath $\gamma $}  \cdot  ({\bf k} - {\bf p})}
  { v_F^2 ({\bf k} - {\bf p})^2 + (\overline{\omega}_k - \overline{\omega}_p)^2 + v_F^2 \nu^2 }   \:
         \frac{1}{{\bf p}^2}  \frac{1}{({\bf p}^2 + \overline{\omega}_p^2/v_F^2)^{n\epsilon/2}}  
                                                                  \nonumber         \\
  & \approx &  v_F \mbox{\boldmath $\gamma $} \cdot {\bf k}  \frac{2}{(4\pi )^{2-\epsilon/2} } 
   \frac{1 + (1-\epsilon) \left(1+\tfrac{n}{2}\epsilon \right) }{(1-\epsilon)(3-\epsilon) }  
 \frac{\Gamma \left(\tfrac{n+1}{2} \epsilon  \right) 
                         \Gamma \left(1-\tfrac{n+1}{2}\epsilon \right)  }
 {\Gamma \left(2 - \tfrac{\epsilon}{2} \right)} \frac{1}{|\nu |^{(n+1)\epsilon }} 
\label{jn}
\end{eqnarray}
 
With the help of (\ref{in}) and (\ref{jn}), one can obtain analytically for
instance the first terms in the expansions of the residues in (\ref{zpn}) and 
(\ref{zvn}), by imposing that the renormalized fermion propagator becomes 
finite in the limit $\epsilon \rightarrow 0$. We find in this way
\begin{eqnarray}
c_1 (g ) & = &  - \frac{1}{24} g^2 - \frac{1}{162} \: g^3 
           -  \frac{5}{5184}    \:  g^4                           
        -  \left( \frac{1}{6480} + \frac{\zeta (3)}{6480} \right) \: g^5  
                                                                \nonumber  \\
  &  &  - \left( \frac{7}{279936}+\frac{\pi ^4}{2799360}+ \frac{\zeta (3)}{34992} \right) \: g^6  
                                                 +  \ldots    \label{cf}     \\
c_2 (g ) & = &  -  \frac{1}{108} \: g^3  - \frac{1}{648}  \: g^4 
                  -  \frac{1}{3888} \: g^5    
         -  \left( \frac{1}{23328}+\frac{\zeta (3)}{23328} \right) \: g^6 
                                                          +  \ldots         \\
c_3 (g ) & = &  - \frac{1}{432} \: g^4 
      - \frac{1}{2430} \: g^5 -  \frac{5}{69984} \: g^6 
                                                           + \ldots          \\
c_4 (g ) & = &   - \frac{1}{1620} \: g^5 
               -  \frac{1}{8748} \: g^6  
                                                          +  \ldots          \\
c_5 (g ) & = &  - \frac{1}{5832} \: g^6 
                                                          +     \ldots         
\label{cl}
\end{eqnarray}
and for the Fermi velocity renormalization
\begin{eqnarray}
b_1 (g ) & = &  - c_1 (g) - \frac{1}{3} g - \frac{1}{72} \: g^2 
           -  \frac{1}{324}    \:  g^3                           
        -  \left( \frac{1}{1728}+\frac{\zeta (3)}{1296} \right) \: g^4  
                                                                \nonumber  \\
  &  &  - \left( \frac{1}{9720}+\frac{\pi ^4}{583200}+\frac{\zeta (3)}{19440} \right) \: g^5  
                                                                \nonumber  \\
  &  &  -  \left(  \frac{5}{279936}+\frac{\pi^4}{8398080}
                 +\frac{\zeta (3)}{69984}+\frac{\zeta (5)}{23328}  \right) \: g^6
                                                   +  \ldots     \label{bf}     \\
b_2 (g ) & = &  - c_2 (g) - \frac{1}{18} \: g^2  - \frac{1}{324}  \: g^3 
                  -  \frac{1}{1296} \: g^4    
         -  \left( \frac{1}{6480}+\frac{\zeta (3)}{4860} \right) \: g^5  
                                                              \nonumber      \\
  &  & - \left( \frac{1}{34992}+\frac{\pi ^4}{2099520}+\frac{\zeta (3)}{69984} \right) \: g^6
                                                          +  \ldots         \\
b_3 (g ) & = &  - c_3 (g) - \frac{1}{81} \: g^3 
      - \frac{1}{1296} \: g^4 -  \frac{1}{4860} \: g^5 
     - \left( \frac{1}{23328}+\frac{\zeta (3)}{17496}  \right) \: g^6
                                                           + \ldots          \\
b_4 (g ) & = &   - c_4 (g) - \frac{1}{324} \: g^4 
               -  \frac{1}{4860} \: g^5  - \frac{1}{17496} \: g^6
                                                          +  \ldots          \\
b_5 (g ) & = &  - c_5 (g) - \frac{1}{1215} \: g^5  - \frac{1}{17496} \: g^6
                                                          +     \ldots       \\
b_6 (g ) & = &   - \frac{1}{4374} \: g^6      +      \ldots
\label{bl}
\end{eqnarray}
The important point is that these expansions of the residues do not show any
dependence on the auxiliary scales $\mu $ and $\nu $ (or on the external 
frequency and momentum when these are used alternatively to regularize the 
self-energy in the infrared). This is a nice check of the 
renormalizability of the theory which guarantees that, at least in the 
large-$N$ approximation, the high-energy divergences can be absorbed into a 
finite number of renormalization factors depending only on the renormalized 
coupling $g$.

The renormalization factors $Z_\psi$ and $Z_v$ provide important information 
about the behavior of the electron system in the low-energy limit. In the 
renormalized theory, the Dirac fermion field gets an anomalous scaling 
dimension $\gamma_\psi (g)$, which can be computed as\cite{amit} 
\begin{equation}
\gamma_\psi = \frac{\mu }{Z_\psi }  \frac{\partial Z_\psi }{\partial \mu }
\end{equation}
This anomalous dimension governs the scaling of the correlators involving 
the Dirac field. When sitting at a fixed-point of the renormalized 
parameters, we would get for instance for the Dirac propagator
\begin{equation}
G(s{\bf k}, s\omega ) \approx  s^{-1 + \gamma_\psi }  G({\bf k}, \omega )
\label{anom}
\end{equation}
On the other hand, the renormalized Fermi velocity $v_R$ gets also a scaling
dependence which can be assessed in terms of a function $\gamma_v (g)$ such 
that
\begin{equation}
\frac{\mu }{v_R }  \frac{\partial v_R}{\partial \mu }  = \gamma_v (g)
\label{vrsc}
\end{equation}
Thus, the knowledge of $\gamma_\psi (g)$ and $\gamma_v (g)$ allows us to 
inspect the theory in the limit of long wavelengths and low energies as 
$\mu \rightarrow 0$.

The anomalous dimension $\gamma_\psi (g)$ can be computed from the dependence of
the renormalized coupling $g$ on the auxiliary scale $\mu $, taking into account 
that
\begin{equation}
\gamma_\psi = \frac{\mu }{Z_\psi }   \frac{\partial g }{\partial \mu }
          \frac{\partial Z_\psi }{\partial g }
\label{chain}
\end{equation}
The scaling of $g$ can be obtained in turn by differentiating the expression
(\ref{eren}) with respect to $\mu $, bearing in mind the independence of $e_0$
with respect to that auxiliary parameter. This leads to 
\begin{equation}
\mu \frac{\partial }{\partial \mu } e^2 = - \epsilon e^2
   +  \frac{N}{6\pi^2 v_F }  e^4
\end{equation}
As already mentioned, the difference between $v_F$ and $v_R$ can be taken 
formally as a quantity of order $\sim 1/N$, so that we end up in the large-$N$ 
limit with the equation
\begin{equation}
\mu \frac{\partial }{\partial \mu } g = - \epsilon g
   +  \frac{1}{3 } g^2
\label{gren}
\end{equation}
This expression can be plugged into (\ref{chain}) to get
\begin{equation}
\gamma_\psi = \frac{1}{Z_\psi } \frac{1}{N} \left( 
  - g \sum_{i=0}^{\infty } \frac{1}{\epsilon^i} \frac{d}{d g} c_{i+1} (g) + \frac{1}{3} g^2 
                \sum_{i=1}^{\infty } \frac{1}{\epsilon^i} \frac{d}{d g} c_{i} (g)  \right)
\label{poles}
\end{equation}

Working to leading order in the $1/N$ expansion, we can set $Z_\psi = 1$ at 
the right-hand-side of (\ref{poles}). The term with no poles in that equation
leads to the result
\begin{equation}
\gamma_\psi = - \frac{1}{N} g \frac{d}{d g}  c_1 (g)
\end{equation}
This expression of the anomalous dimension makes only sense, however, provided
that one can certify the cancellation of the rest of terms carrying poles of 
all orders in the $\epsilon $ parameter, which implies the hierarchy of 
equations
\begin{equation}
\frac{d}{d g} c_{i+1} (g)  - \frac{1}{3} g \frac{d}{d g} c_i (g) = 0
\label{hi}
\end{equation}

It is remarkable that the power series of the residues given in 
(\ref{cf})-(\ref{cl}) satisfy indeed the hierarchy (\ref{hi}). Beyond the 
analytic approach, we have computed numerically the expansion of the 
first residues $c_i (g)$ up to order $g^{32}$, with a precision of 40 digits. 
In this way, we have been able to check that the conditions (\ref{hi}) are 
fulfilled at that level of approximation, reassuring the 
consistency of the large-$N$ approach for the present field theory.

\begin{figure}

\vspace{0.5cm}

\begin{center}
\mbox{\epsfxsize 7cm \epsfbox{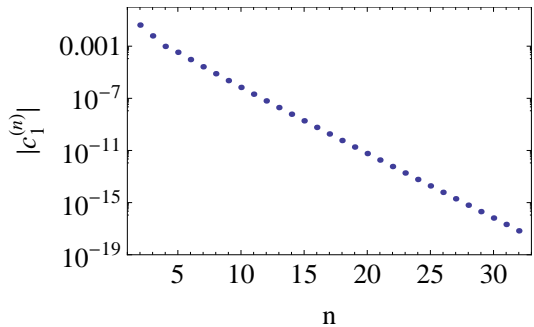}} 

\end{center}
\caption{Plot of the absolute value of the coefficients $c_1^{(n)}$ in the 
expansion of $c_1 (g)$ as a power series of the renormalized coupling $g$.}
\label{six}
\end{figure}

Another important detail of the numerical calculation of the residues is the
evidence that the coefficients of each perturbative expansion approach a 
geometric sequence for large orders of the coupling. In the case of the residue 
$c_1 (g)$, we have for instance the power series
\begin{equation}
c_1 (g )  = \sum_{n=1}^{\infty}  c_1^{(n)} g^n
\label{geom}
\end{equation}
with coefficients that we have represented in Fig. \ref{six} up to the order
we have carried out the numerical computation. The behavior observed for the 
series $c_1^{(n)}$ implies that the perturbative expansion must have a finite
radius of convergence. This can be obtained by approaching the ratio between
consecutive coefficients according to the dependence
\begin{equation}
\frac{c_1^{(n)}}{c_1^{(n-1)}} = 
          r + \frac{r'}{n} + \frac{r''}{n^2} + \frac{r'''}{n^3} + \ldots
\end{equation}
which provides indeed an excellent fit at large $n$. We get in this way the 
estimate
\begin{equation}
r \approx 0.3333333
\label{critn}
\end{equation}
where the error lies in the last digit. We find then the singular point for
the coupling $g$ (the radius of convergence) at 
\begin{equation}
g_c = \frac{1}{r} \approx 3.0 \pm 1.0 \times 10^{-7}
\label{gc3}
\end{equation}

\begin{figure}

\vspace{0.5cm}

\begin{center}
\mbox{\epsfxsize 7cm \epsfbox{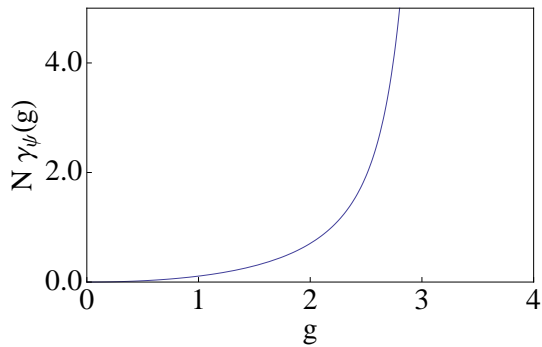}} 

\end{center}
\caption{Plot of the anomalous scaling dimension $\gamma_\psi (g)$, multiplied by
the number $N$ of four-component Dirac fermions in the electron system.}
\label{seven}
\end{figure}

The critical value $g_c$ corresponds to a point where the anomalous scaling 
dimension $\gamma_\psi (g)$ diverges. This can be appreciated in Fig. 
\ref{seven}, where we have represented that function from the results of our 
numerical calculation. The singularity found at $g_c$ has a precise physical 
meaning, since $\gamma_\psi $ governs the scaling of all the correlators 
involving the Dirac fermion field. Away from a fixed-point in the renormalized 
parameters, the scaling is not so simple as in (\ref{anom}), but from that 
expression we already get the idea that the divergence of $\gamma_\psi$ implies 
the decay of the Dirac propagator\cite{barnes2}. In the limit 
$g \rightarrow g_c$, the singularity in the scaling dimension leads indeed to 
the suppression of the quasiparticle weight. This characterizes a particular 
form of correlated behavior, which has been identified in several other 
instances of interacting electrons and has led to constitute the class of 
so-called non-Fermi liquids\cite{bares,nayak,hou,cast}. The distinctive 
feature of this class is the absence of a quasiparticle pole in the electron 
propagator, which leads to a very appealing paradigm to explain unconventional 
properties like those found in the normal state of copper-oxide 
superconductors.

In order to ensure the relevance of the divergence at $g_c$, we have anyhow
to verify that such a singular behavior is not prevented by some other 
instability in the low-energy scaling of the electron system. We pay attention
in particular to the renormalization of the Fermi velocity, which has a natural
tendency to grow in the low-energy limit. The function $\gamma_v (g)$ that 
dictates the scaling in Eq. (\ref{vrsc}) can be obtained from the independence 
of the bare Fermi velocity on the auxiliary scale $\mu $:
\begin{equation}
\frac{\mu }{v_R} \frac{\partial }{\partial \mu }  \left(  Z_v v_R  \right)  =  0
\label{sfree}
\end{equation}
The dependence of $Z_v$ on $\mu $ comes only from the coupling $g$ so that, 
using Eq. (\ref{gren}), we get
\begin{equation}
Z_v \frac{\mu }{v_R} \frac{\partial }{\partial \mu } v_R 
   - \frac{1}{N} g \sum_{i=0}^{\infty } 
                  \frac{1}{\epsilon^i} \frac{d}{d g} b_{i+1} (g)
  + \frac{1}{N} \frac{1}{3} g^2 \sum_{i=1}^{\infty } 
                    \frac{1}{\epsilon^i} \frac{d}{d g} b_{i} (g)  = 0
\label{poles2}
\end{equation}
Working to leading order in the $1/N$ expansion, the term free of poles in 
Eq. (\ref{poles2}) leads to the result
\begin{equation}
\frac{\mu }{v_R} \frac{\partial }{\partial \mu } v_R = \frac{1}{N} g \frac{d}{d g} b_1 (g)
\end{equation}
As in the case of the anomalous scaling dimension, one has to make sure 
however that the rest of terms carrying poles of all orders in 
Eq. (\ref{poles2}) cancel out identically, which is enforced by the conditions
\begin{equation}
\frac{d}{d g} b_{i+1} (g)  - \frac{1}{3} g \frac{d}{d g} b_i (g) = 0
\label{hip}
\end{equation}

\begin{figure}

\vspace{0.5cm}

\begin{center}
\mbox{\epsfxsize 7cm \epsfbox{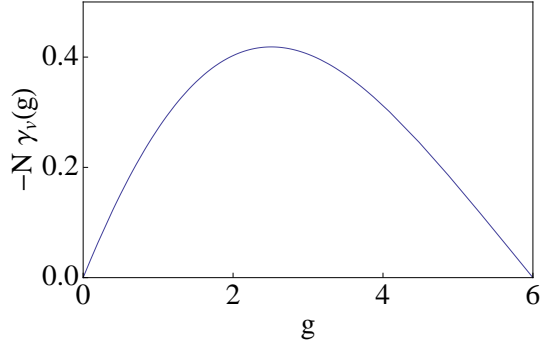}} 

\end{center}
\caption{Plot of the rate of variation $\gamma_v (g)$ of the renormalized Fermi 
velocity with respect to energy, multiplied by the number $N$ of four-component 
Dirac fermions in the electron system.}
\label{eight}
\end{figure}

The expressions of the residues in (\ref{bf})-(\ref{bl}) satisfy indeed the set 
of equations (\ref{hip}), and a more extensive numerical computation confirms 
that this is also the case for the power series of the $b_i (g)$ evaluated up 
to order $g^{32}$. This analysis also shows that these expansions do not lead 
to any singularity in the range of couplings up to the critical point $g_c$. 
The corresponding function $\gamma_v (g)$ obtained from $b_1 (g)$ is 
represented in Fig. \ref{eight}. The regular behavior observed in the plot 
implies that the scaling of the Fermi velocity is not an obstacle for the 
suppression of the fermion quasiparticles as the interaction strength becomes 
sufficiently large to hit the critical point at $g_c$.

\begin{figure}

\vspace{0.5cm}

\begin{center}
\mbox{\epsfxsize 4cm \epsfbox{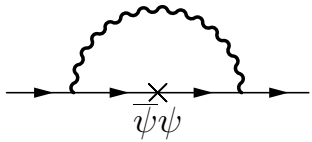}} 

\end{center}
\caption{Corrections to the vertex for the mass operator in the large-$N$ 
approximation. The cross represents the operator $\overline{\psi }  \psi $ and 
the thick wiggly line stands for the dressed interaction propagator as defined 
in Fig. \ref{five}.}
\label{nine}
\end{figure}

As a last check, we look also for the possible tendency towards chiral symmetry 
breaking and dynamical mass generation in the large-$N$ approximation. For this 
purpose, we may analyze the scaling of the vertex for the mass operator 
represented in Fig. \ref{nine}. Computing in the limit where both frequency and 
momentum transfer vanish, we get
\begin{eqnarray}
\Gamma_m ({\bf 0}, 0;{\bf k}, \omega_k)  & = &  
  1  +  i  \int \frac{d^D p}{(2\pi )^D} \frac{d\omega_p }{2\pi } \:  \gamma_0
 \left( -\frac{\gamma_0 (\omega_k - \omega_p ) + 
             v_F \mbox{\boldmath $\gamma $}  \cdot  ({\bf k} - {\bf p})}
  { - (\omega_k - \omega_p)^2 + v_F^2 ({\bf k} - {\bf p})^2 - i\eta}  \right)^2  \gamma_0 \:
                                                         \nonumber           \\
     &    &    \;\;\;\;\;\;\;\;\;\;\;\;\;\;\;\;\;\;\;\;\;\;\;\;\;   
                  \times   \frac{e_0^2}
             {{\bf p}^2 \left(1 + N C(\epsilon ) \frac{e_0^2}{2\pi^2 v_F}
       \frac{1}{({\bf p}^2 - \omega_p^2/v_F^2)^{\epsilon/2}} \right)}
\end{eqnarray}
We have to account as before for the renormalization of the charge, which leads 
to an expression in terms of the renormalized coupling $g$
\begin{eqnarray}
\Gamma_m ({\bf 0}, 0;{\bf k}, i\overline{\omega}_k)  & = &  
  1  +     \frac{2\pi^2 v_F}{N}  \mu^\epsilon \int \frac{d^D p}{(2\pi )^D} 
                         \frac{d\overline{\omega}_p }{2\pi }   \:
  \frac{1}{ v_F^2 ({\bf k} - {\bf p})^2 + (\overline{\omega}_k - \overline{\omega}_p)^2}   \:
         \frac{1}{{\bf p}^2}                             \nonumber       \\
    &   &      \;\;\;\;\;\;\;\;\;\;\;\;\;\;\;\;\;\;\;\;\;\;\;\;\; 
      \sum_{n=0}^{\infty} (-1)^n g^{n+1} \left(- \frac{1}{3\epsilon } + C(\epsilon )
       \frac{\mu^\epsilon}{({\bf p}^2 + \overline{\omega}_p^2/v_F^2)^{\epsilon/2}} \right)^n 
\label{vertn}
\end{eqnarray}

As already done in the case of the electron self-energy, we can compute the 
high-energy divergences of the vertex in the limit of vanishing ${\bf k}$ and 
$\overline{\omega}_k$, regularizing the integrals in the infrared with an 
auxiliary mass $\nu $ in the Dirac propagators. In this way, the different terms in the expansion 
(\ref{vertn}) can be obtained using the general result
\begin{eqnarray}     
K_n  & = & v_F \int \frac{d^D p}{(2\pi )^D} \frac{d\overline{\omega}_p }{2\pi }   \:
  \frac{1}
  { v_F^2 {\bf p}^2 +  \overline{\omega}_p^2 + v_F^2 \nu^2 }   \:
         \frac{1}{{\bf p}^2}  \frac{1}{({\bf p}^2 + \overline{\omega}_p^2/v_F^2)^{n\epsilon/2}}  
                                                                  \nonumber         \\
  & = &   \frac{1}{(4\pi )^{2-\epsilon/2} } 
   \frac{2}{1-\epsilon }  
 \frac{\Gamma \left(\tfrac{n+1}{2} \epsilon  \right) 
                         \Gamma \left(1-\tfrac{n+1}{2}\epsilon \right)  }
 {\Gamma \left(1 - \tfrac{\epsilon}{2} \right)} \frac{1}{|\nu |^{(n+1)\epsilon }} 
\label{kn}
\end{eqnarray}

We see that the vertex can develop in general divergences of order 
$\sim 1/\epsilon^n $ at the level $g^n$ in the perturbative expansion. These 
have to be reabsorbed into a multiplicative renormalization of the type shown
in (\ref{zm}). The point to bear in mind is that now $Z_m$ is made of two
different factors, coming independently from the renormalization of the 
composite mass operator (\ref{mass}) and the Dirac fermion fields in the  
vertex\cite{amit}:
\begin{equation}
Z_m  =    Z_{\psi } Z_{\psi^2 }       
\end{equation}
The renormalization factor $Z_{\psi^2 }$ for the mass operator can have the 
general structure
\begin{equation}
Z_{\psi^2 } = 1 + \frac{1}{N}\sum_{i=1}^{\infty } \frac{\bar{d}_i (g)}{\epsilon^i} 
                                      +  O \left( \frac{1}{N^2} \right)
\end{equation}
Under the assumption that the theory is renormalizable, it must be possible 
to choose appropriately the residues $\bar{d}_i (g)$ to end up with a 
renormalized vertex, finite in the limit $\epsilon \rightarrow 0$, given by
\begin{equation}
\Gamma_{m,{\rm ren}} =  Z_\psi Z_{\psi^2 }  \: \Gamma_m
\label{gmr}
\end{equation}

We have checked that the vertex (\ref{gmr}) can be made free of poles, at least 
up to order $g^{32}$ we have been able to compute it numerically, with a set of
residues $\bar{d}_i (g)$ that depend only on the renormalized coupling $g$. We 
have also seen that these functions have a regular behavior in the range
up to the critical coupling $g_c$ where the anomalous dimension $\gamma_\psi$
diverges. This makes clear that the dominant instability in the large-$N$ 
approximation is indeed characterized by the suppression of the electron 
quasiparticles.

\begin{figure}[h]

\vspace{0.5cm}

\begin{center}
\mbox{\epsfxsize 7cm \epsfbox{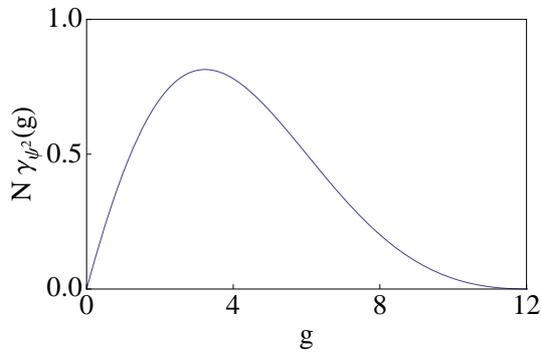}} 

\end{center}
\caption{Plot of the anomalous scaling dimension $\gamma_{\psi^2 } (g)$, 
multiplied by the number $N$ of four-component Dirac fermions in the electron 
system.}
\label{ten}
\end{figure}

Of all the residues $\bar{d}_i (g)$, the first of them conveys the most 
relevant piece of information, since it is related to the anomalous dimension
of the composite mass operator, defined by
\begin{equation}
\gamma_{\psi^2 } = \frac{\mu }{Z_{\psi^2 } }  
                 \frac{\partial Z_{\psi^2 } }{\partial \mu }
\label{ad2}
\end{equation}
Paralleling the above derivation of similar scaling dimensions, one arrives at 
the result
\begin{equation}
\gamma_{\psi^2 } = - \frac{1}{N} g \frac{d}{d g} \bar{d}_1 (g)
\end{equation}
The plot of this function obtained from our numerical computation of 
$\bar{d}_1 (g)$ is represented in Fig. \ref{ten}. The regular behavior
observed in the figure is the evidence that no singularity can be expected
in the correlators of the mass operator $\rho_m $, whose magnitude has got
to be bound to the scaling dictated by the anomalous dimension 
$\gamma_{\psi^2 }$ in the low-energy limit.

\section{Conclusions}

We have studied the development of the strong-coupling phases in the QED of 
3D Dirac and Weyl semimetals by means of two different nonperturbative 
approaches, consisting in the sum of ladder diagrams on one hand, and
taking the limit of large number of fermion flavors on the other hand. 
We have benefited from the renormalizability that the theory shows in both 
cases, which makes possible to render all the renormalized quantities 
independent of the high-energy cutoff. Thus we have been able to compute the 
anomalous scaling dimensions of a number of operators exclusively in terms of 
the renormalized coupling constant, which has allowed us to determine the 
precise location of the singularities signaling the onset of the 
strong-coupling phases.

We have seen that, in the ladder approximation, the 3D Dirac semimetals have 
a critical point marking the boundary of a phase with chiral symmetry breaking 
and dynamical generation of mass, in analogy with the same strong-coupling 
phenomenon in conventional QED\cite{mas,fom,fuk,mir,gus2,kon,atk,min}. 
We have found however that such a breakdown of
symmetry does not persist in the large-$N$ limit of the theory, which is 
instead characterized by the growth of the anomalous dimension of the electron
quasiparticles at large interaction strength. The picture that emerges by 
combining the results from the two nonperturbative approaches is that chiral
symmetry breaking must govern the strong-coupling physics of the 3D Dirac  
semimetals for sufficiently small $N$, while there has to be a transition to a 
different phase with strong suppression of electron quasiparticles prevailing
above a certain value of $N$.

\begin{figure}[h]
\begin{center}
\includegraphics[width=6.0cm]{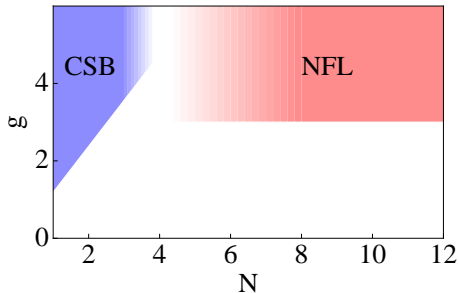}
\end{center}
\caption{Phase diagram of the QED of 3D Dirac semimetals showing an approximate
map of the phases corresponding to chiral symmetry breaking (CSB) and non-Fermi
liquid behavior (NFL), obtained from the values of the respective critical 
couplings in the ladder approach and the large-$N$ approximation.}
\label{eleven}
\end{figure}

With the results obtained for the different critical points we can draw 
an approximate phase diagram in terms of the number $N$ of Dirac fermions and
the renormalized coupling $g$ defined in (\ref{gdef}). In principle, the 
critical coupling (\ref{gc3}) can provide a reliable estimate for the onset of 
non-Fermi liquid behavior at sufficiently large $N$. On the other hand, the 
critical value of $g$ deriving from (\ref{crit2}) may lead to a sensible map of 
the phase with chiral symmetry breaking as long as its magnitude does not 
become larger than that of (\ref{gc3}). The resulting phase diagram of the 3D 
Dirac semimetals, represented in Fig. \ref{eleven}, shows the regions where the 
behavior of the phase boundaries may be captured by our alternative 
approximations, away from the intermediate regime about $N = 4$ where the 
competition between the two strong-coupling phases cannot be reliably described 
within our analytic framework.

It is interesting to compare at this point the phase diagram in Fig. 
\ref{eleven} with that obtained with the resolution of the Schwinger-Dyson
equations, which can be trusted for all values of $N$. We can see from the
results of Ref. \cite{qed} that such a numerical approach sets indeed the 
interplay between the phases with chiral symmetry breaking and non-Fermi liquid 
behavior at $N = 4$. For $N > 4$, we observe that the critical line for that 
latter phase is not straight, although it approaches a 
constant asymptotic limit at large $N$. The self-consistent resolution leads 
also to a critical line for chiral symmetry breaking in the regime $N \leq 4$ 
that has an approximate linear behavior as a function of $N$. It is worth to 
mention that the values of the critical couplings found in the numerical 
resolution of Ref. \cite{qed} are at first sight much larger than their 
counterparts in the diagram of Fig. \ref{eleven}. We have to bear in mind, 
however, that the critical values in that paper were given for the bare 
couplings, in the theory with a finite high-energy cutoff, while critical 
points like (\ref{crit2}) and (\ref{gc3}) are referred in the present context 
to the renormalized couplings. Overall, we may conclude that there is good 
qualitative agreement between the location of the phases in the diagram of 
Fig. \ref{eleven} and in the more complete map obtained from the 
resolution of the Schwinger-Dyson equations.

We end up remarking that our analysis can be applied to characterize 
%equally well 
not only the strong-coupling regime of the 3D Dirac semimetals, but also of
the Weyl semimetals. In these materials, each Weyl point hosts a fermion with 
a definite chirality, which means that the electron system cannot undergo 
chiral symmetry breaking through condensation of some order parameter at 
zero momentum\cite{hut}. This obstruction does not hold however for the 
strong-coupling phase that we have identified in terms of the suppression of 
fermion quasiparticles. It is clear that the large-$N$ approach of Sec. IV 
applies equally well for Weyl and Dirac fermions, so that Weyl semimetals 
are susceptible of developing the phase with non-Fermi liquid behavior that 
we have mapped at large $N$ in the diagram of Fig. \ref{eleven}. In this 
respect, there should be good prospects to observe such an unconventional 
behavior in present candidates for Weyl semimetals, like TaAs or the pyrochlore 
iridates, which have up to 12 pairs of Weyl points\cite{taas1,taas2}. 
These considerations show that the strong-coupling phases studied here are not 
beyond reach, and that they may be actually realized in 3D Dirac or Weyl 
semimetals with suitably small values of the Fermi velocity or with the large 
number of fermion flavors already exhibited by known materials.

\section{Acknowledgments}

We acknowledge the financial support from MICINN (Spain) through grant 
FIS2011-23713 and from MINECO (Spain) through grant FIS2014-57432-P.




\vspace{2cm}



\begin{thebibliography}{99}



\bibitem{novo}
K. S. Novoselov, A. K. Geim, S. V. Morozov, D. Jiang, Y. Zhang, S. V. Dubonos, 
I. V. Grigorieva, and A. A. Firsov, Science {\bf 306} (2004) 666.

\bibitem{liu}
Z. K. Liu, B. Zhou, Y. Zhang, Z. J. Wang, H. M. Weng, D. Prabhakaran, S.-K. Mo, Z. X. Shen, Z. Fang, X. Dai, Z. Hussain and Y. L. Chen, Science {\bf 343}, 864 (2014).

\bibitem{neu}
M. Neupane, S. Xu, R. Sankar, N. Alidoust, G. Bian, C. Liu, I. Belopolski, T.-R. Chang, H.-T. Jeng, H. Lin, A. Bansil, F. Chou and M. Z. Hasan, Nature Commun. {\bf 5}, 3786 (2014).

\bibitem{bor}
S. Borisenko, Q. Gibson, D. Evtushinsky, V. Zabolotnyy, B. B\"uchner and R. J. Cava, Phys. Rev. Lett. {\bf 113}, 027603 (2014).

\bibitem{taas1}
S.-M. Huang, S.-Y. Xu, I. Belopolski, C.-C. Lee, G. Chang, B. Wang, N. Alidoust, G. Bian, M. Neupane, C. Zhang, S. Jia, A. Bansil, H. Lin and M. Z. Hasan, Nature Commun. {\bf 6}, 7373 (2015).

\bibitem{taas2}
S.-Y. Xu, I. Belopolski, N. Alidoust, M. Neupane, C. Zhang, R. Sankar, S.-M. Huang, C.-C. Lee, G. Chang, B. Wang, G. Bian, H. Zheng, D. S. Sanchez, F. Chou, H. Lin, S. Jia and M. Z. Hasan, report arXiv:1502.03807 (to appear in Science).

\bibitem{kats}
M. I. Katsnelson, K. S. Novoselov and A. K. Geim, Nature Phys. {\bf 2} (2006) 620.

\bibitem{ando}
H. Suzuura and T. Ando, Phys. Rev. Lett. {\bf 89} (2002) 266603.

\bibitem{kane}
M. Z. Hasan and C. L. Kane, Rev. Mod. Phys. 82, 3045 (2010).

\bibitem{qi}
X.-L. Qi and S.-C. Zhang, Physics Today 63, 33 (2010).

\bibitem{nil}
V. M. Pereira, J. Nilsson and A. H. Castro Neto, Phys. Rev. Lett. {\bf 99} (2007) 166802.

\bibitem{fog}
M. M. Fogler, D. S. Novikov, and B. I. Shklovskii, Phys. Rev. B {\bf 76} (2007) 233402.

\bibitem{shy}
A. V. Shytov, M. I. Katsnelson, and L. S. Levitov, 
Phys. Rev. Lett. {\bf 99} (2007) 236801.

\bibitem{ter}
I. S. Terekhov, A. I. Milstein, V. N. Kotov, and O. P. Sushkov,
Phys. Rev. Lett. {\bf 100} (2008) 076803.

\bibitem{np2}
J. Gonz\'alez, F. Guinea and M. A. H. Vozmediano,
Nucl. Phys. B {\bf 424} (1994) 595.

\bibitem{prbr}
J. Gonz\'alez, F. Guinea and M. A. H. Vozmediano,
Phys. Rev. B {\bf 59} (1999) R2474.

\bibitem{exp2}
D. C. Elias, R. V. Gorbachev, A. S. Mayorov, S. V. Morozov, A. A. Zhukov, P. Blake,
L. A. Ponomarenko, I. V. Grigorieva, K. S. Novoselov, F. Guinea and A. K. Geim,
Nature Phys. {\bf 7} (2011) 701.

\bibitem{jhep}
J. Gonz\'alez, JHEP {\bf 08}, 27 (2012).

\bibitem{gama}
O. V. Gamayun, E. V. Gorbar and V. P. Gusynin, Phys. Rev. B {\bf 80} (2009) 165429.

\bibitem{mas}
T. Maskawa and H. Nakajima, Prog. Theor. Phys. {\bf 52}, 1326 (1974).

\bibitem{fom}
P. I. Fomin and V. A. Miransky, Phys. Lett. {\bf 64B}, 166 (1976).

\bibitem{fuk}
R. Fukuda and T. Kugo, Nucl. Phys. B {\bf 117}, 250 (1976).

\bibitem{mir}
V. A. Miransky, Nuovo Cimento {\bf 90A}, 149 (1985); Sov. Phys. JETP {\bf 61}, 
905 (1985).

\bibitem{gus2}
V. P. Gusynin, Mod. Phys. Lett. A {\bf 5}, 133 (1990).

\bibitem{kon}
K.-I. Kondo and H. Nakatani, Nucl. Phys. B  {\bf 351}, 236 (1991).

\bibitem{atk}
D. Atkinson, H. J. De Groot and P. W. Johnson, Int. J. Mod. Phys. A {\bf 7}, 
7629 (1992).

\bibitem{min}
K.-I. Kondo, H. Mino and H. Nakatani, Mod. Phys. Lett. A {\bf 7}, 
1509 (1992).

\bibitem{rc}
J. Gonz\'alez, Phys. Rev. B {\bf 90}, 121107(R) (2014).

\bibitem{ram}
P. Ramond, {\em Field Theory: A Modern Primer}, Benjamin/Cummings, Reading (1981).

\bibitem{ros}
This scaling of the Fermi velocity in 3D Dirac semimetals has been pointed 
out by P. Hosur, S. A. Parameswaran, and A. Vishwanath, Phys. Rev. Lett. 
{\bf 108}, 046602 (2012), and also by B. Rosenstein and M. Lewkowicz, 
Phys. Rev. B {\bf 88}, 045108 (2013).

\bibitem{amit}
D. J. Amit and V. Mart\'{\i}n-Mayor, {\em Field Theory, the Renormalization 
Group, and Critical Phenomena}, World Scientific, Singapore (2005).

\bibitem{nomu}
This phase has been also studied in 3D Dirac semimetals by
A. Sekine and K. Nomura, Phys. Rev. B {\bf 90}, 075137 (2014).

\bibitem{sabio}
J. Sabio, F. Sols and F. Guinea, Phys. Rev. B {\bf 82}, 121413(R) (2010).

\bibitem{qed}
J. Gonz\'alez, report arXiv:1502.07640 (to appear in Phys. Rev. B).

\bibitem{landau}
E. M. Lifshitz and L. P. Pitaevskii, {\em Relativistic Quantum Theory (Course of Theoretical Physics Volume 4, Part 2)}, Pergamon Press, Oxford (1974).

\bibitem{barnes2}
Signatures of the breakdown of the Fermi liquid picture in the 3D Dirac 
semimetals have been also found by J. Hofmann, E. Barnes and S. Das Sarma,
Phys. Rev. B {\bf 92}, 045104 (2015). 

\bibitem{bares}
P.-A. Bares and X. G. Wen, Phys. Rev. B {\bf 48}, 8636 (1993).

\bibitem{nayak}
C. Nayak and F. Wilczek, Nucl. Phys. B {\bf 417}, 359 (1994).

\bibitem{hou}
A. Houghton, H.-J. Kwon, J. B. Marston and R. Shankar, 
J. Phys.: Condens. Matter {\bf 6}, 4909 (1994).

\bibitem{cast}
C. Castellani, S. Caprara, C. Di Castro and A. Maccarone,
Nucl. Phys. B {\bf 594}, 747 (2001).

\bibitem{hut}
One cannot discard however the possible development of a condensate with 
nonvanishing momentum, as analyzed by R.-X. Zhang, J. A. Hutasoit, Y. Sun, 
B. Yan, C. Xu, and C.-X. Liu, report arXiv:1503.00358.






\end{thebibliography}
\end{document}